\documentclass[sigconf]{acmart}
\AtBeginDocument{%
  }

\copyrightyear{2025}
\acmYear{2025}
\setcopyright{cc}
\setcctype{BY}
\acmConference[KDD '25]{Proceedings of the 31st ACM SIGKDD Conference on Knowledge Discovery and Data Mining V.1}{August 3--7, 2025}{Toronto, ON, Canada}
\acmBooktitle{Proceedings of the 31st ACM SIGKDD Conference on Knowledge Discovery and Data Mining V.1 (KDD '25), August 3--7, 2025, Toronto, ON, Canada}
\acmDOI{10.1145/3690624.3709278}
\acmISBN{979-8-4007-1245-6/25/08}

\usepackage{diagbox}

\usepackage{float}
\usepackage{array}
\usepackage{multirow}
\usepackage{threeparttable}
\usepackage{subfigure}
\usepackage{bbding}
\usepackage{enumitem}
\usepackage{makecell}
\usepackage{soul}
\usepackage{balance}
\setlist[itemize]{leftmargin=*}

\begin{document}

\title{Combinatorial Optimization Perspective based Framework for Multi-behavior Recommendation}

\author{Chenhao Zhai}
\authornote{Both authors contributed equally to this research.}
\authornote{Work done when he was a research intern at Kuaishou Technology.}
\affiliation{
  \institution{Shenzhen International Graduate School, Tsinghua University}
  \city{Shenzhen}
  \country{China}
}
\email{dch23@mails.tsinghua.edu.cn}

\author{Chang Meng}
\authornotemark[1]
\affiliation{
  \institution{Kuaishou Technology}
  \city{Beijing}
  \country{China}
}
\email{mengchang@kuaishou.com}

\author{Yu Yang}
\affiliation{
  \institution{Shenzhen International Graduate School, Tsinghua University}
  \city{Shenzhen}
  \country{China}
}
\email{yu-yang23@mails.tsinghua.edu.cn}

\author{Kexin Zhang}
\affiliation{
  \institution{Shenzhen International Graduate School, Tsinghua University}
  \city{Shenzhen}
  \country{China}
}
\email{zkx21@mails.tsinghua.edu.cn}

\author{Xuhao Zhao}
\affiliation{
  \institution{School of Electronic Information and Electrical Engineering, Shanghai Jiao Tong University}
  \city{Shanghai}
  \country{China}
}
\email{zhaoxuhao@sjtu.edu.cn}

\author{Xiu Li}
\authornote{The corresponding author.}
\affiliation{
  \institution{Shenzhen International Graduate School, Tsinghua University}
  \city{Shenzhen}
  \country{China}
}
\email{li.xiu@sz.tsinghua.edu.cn}

\renewcommand{\shortauthors}{Chenhao Zhai et al.}

\begin{abstract}
   In real-world recommendation scenarios, users engage with items through various types of behaviors. Leveraging diversified user behavior information for learning can enhance the recommendation of target behaviors (e.g., buy), as demonstrated by recent multi-behavior methods. The mainstream multi-behavior recommendation framework consists of two steps: fusion and prediction. Recent approaches utilize graph neural networks for multi-behavior fusion and employ multi-task learning paradigms for joint optimization in the prediction step, achieving significant success. However, these methods have limited perspectives on multi-behavior fusion, which leads to inaccurate capture of user behavior patterns in the fusion step. Moreover, when using multi-task learning for prediction, the relationship between the target task and auxiliary tasks is not sufficiently coordinated, resulting in negative information transfer. To address these problems, we propose a novel multi-behavior recommendation framework based on the combinatorial optimization perspective, named COPF. Specifically, we treat multi-behavior fusion as a combinatorial optimization problem, imposing different constraints at various stages of each behavior to restrict the solution space, thus significantly enhancing fusion efficiency (COGCN). In the prediction step, we improve both forward and backward propagation during the generation and aggregation of multiple experts to mitigate negative transfer caused by differences in both feature and label distributions (DFME). Comprehensive experiments on three real-world datasets indicate the superiority of COPF. Further analyses also validate the effectiveness of the COGCN and DFME modules. Our code is available at \url{https://github.com/1918190/COPF}.
\end{abstract}




\maketitle
\section{INTRODUCTION}
\label{intro}
Recommender system is a crucial technology in today's society \cite{meng2024coarse, wang2024future,li2024cdrnp,zhao2023task,dang2024augmenting}, delivering high-quality personalized recommendations based on user preferences. Delving into users’ historical engagements, traditional collaborative filtering (CF) techniques \cite{cfsurvey} learn the representations of users and items for improved recommendations. 
Although somewhat effective, these methods only account for a single type of user-item interaction, limiting their practical effectiveness. In the real world, user behavior is diverse. Beyond target behavior (e.g., \textit{buy}), which is the primary focus for businesses and platforms, users also engage in behaviors such as viewing, adding to cart, and collecting. This behavioral information encompasses different dimensions of user preferences that can be leveraged as auxiliary knowledge to enhance the learning of target behaviors and provide better service for users \cite{nmtr,matn,mbgcn}.

To fully exploit auxiliary behaviors, multi-behavior recommendation methods have emerged as solutions. These methods can be divided into two steps: multi-behavior fusion and multi-behavior prediction \cite{he2023survey, pkef}. In the fusion step, multiple behaviors are combined to learn the representations that capture user preferences. In the prediction step, these representations are applied for model prediction, with multi-task learning (MTL) proving effective \cite{huang2021recent}.

\begin{figure}[t]
	\centering
	\setlength{\belowcaptionskip}{0cm}
	\setlength{\abovecaptionskip}{0cm}
	\includegraphics[width=0.49\textwidth]{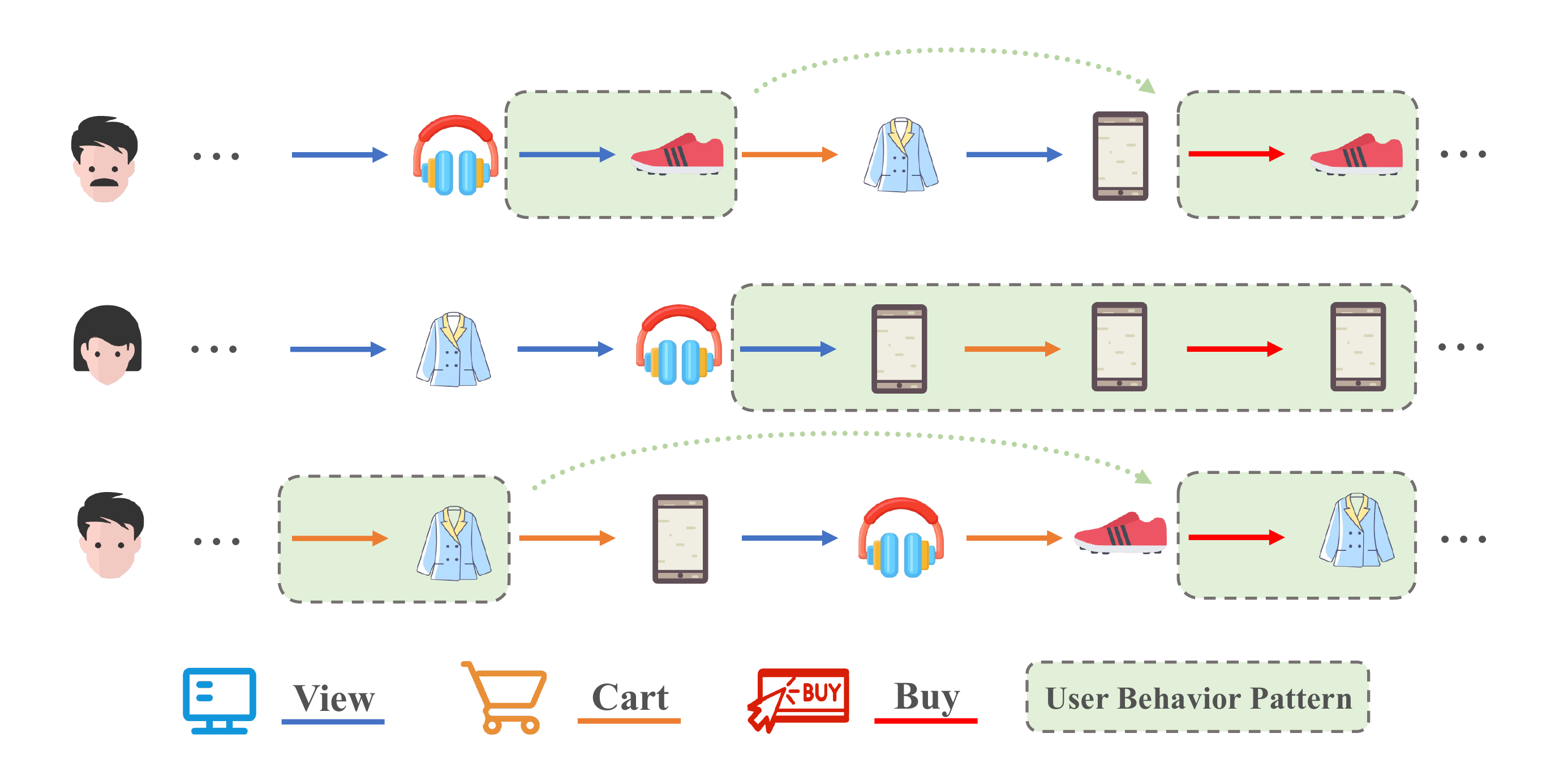}
	\caption{Examples of the user behavior patterns.}
	\label{fig:behavior_pattern}
	\vspace{-4mm}
\end{figure}

With the explosive development of deep learning, most methods employed for multi-behavior fusion have transitioned from traditional matrix factorization \cite{mf1,mf2,mf3} to deep neural networks \cite{matn,nmtr,dipn}. Among them, graph neural networks (GNNs) \cite{lightgcn,ngcf,lr-gccf,hpmr,ckml} have become popular techniques in multi-behavior recommendation due to their ability to model higher-order user-item interactions effectively \cite{khgt,ghcf,gnmr,pkef}. For example, MBGCN \cite{mbgcn} and CIGF \cite{cigf} learn different behavioral information synchronously and fuse them through learnable parameters. Additionally, some studies \cite{crgcn, mbcgcn, pkef} leverage the cascading dependencies between behaviors in the real world (e.g., \textit{view → cart → buy}) to enhance model learning. BCIPM \cite{bipn} emphasizes the target behavior by strategically modeling different behaviors. These above methods essentially share the same goal: \textit{capturing richer user preferences by deeply exploring the complex user behavior patterns formed by heterogeneous interactions (shown in Figure \ref{fig:behavior_pattern})}. However, previous methods either simply aggregate the behavior representations without constraints, or define strict sequential relationships for the behaviors, resulting in inadequate modeling of user behavior patterns.

In the multi-behavior prediction step, MTL modules are widely utilized \cite{mbgmn, CML, crgcn, bipn} to overcome the limitation of single label \cite{mbgcn, smbrec} in representing diverse user preferences. These modules incorporate auxiliary behavior labels for joint optimization, allowing the model to leverage multi-behavior information for improved accuracy. To better adapt to multi-behavior prediction tasks, CIGF \cite{cigf} enhances traditional MTL models \cite{mmoe, PLE} by further decoupling the inputs of different tasks to mitigate gradient conflicts. PKEF \cite{pkef} builds on this by introducing a projection mechanism during aggregation, preventing the incorporation of harmful information.

Although the above multi-behavior recommendation methods have demonstrated effectiveness in multi-behavior fusion and prediction steps respectively, the following challenges still exist:

\begin{itemize}
\item \textbf{Combination optimization for Multi-Behavior Fusion.} Combination optimization problems typically involve numerous states and choices, necessitating an optimal solution within manageable complexity. Multi-behavior recommendation is essentially a combination optimization problem. For a given behavioral category, each user has a finite number of possible user behavior patterns. The optimal behavioral pattern can be yielded by exhaustively enumerating all possibilities, but this results in significant spatial and temporal costs, known as "combinatorial explosion." Therefore, the challenge lies in leveraging existing knowledge to restrict the solution space and achieve efficient multi-behavior recommendations. Early approaches \cite{lightgcn, smbrec} focused solely on learning behavior-specific information without considering user behavior patterns, which was improved by methods \cite{mbgcn, cigf, bipn} that incorporated behavior aggregation during the learning process. However, they still lacked adequate constraints on user behavior patterns. Some recent methods \cite{crgcn, mbcgcn, pkef} use a cascading paradigm to model user behavioral sequences, resulting in overly strict constraints on user behavior patterns. In conclusion, a paradigm that establishes appropriate constraints to restrict the solution space urgently needs to be proposed.

\item \textbf{Coordination of Correlations between Tasks.} Multi-task learning (MTL) is a commonly used approach for multi-behavior prediction, which models different behaviors as independent tasks. Since each task can affect the final prediction, it is crucial to properly coordinate the correlations between tasks, which is currently limited by two factors: \textbf{1) \textit{Differences in feature space distribution during forward propagation}}: Most existing methods overlook this aspect, leading to biases in information aggregation. While recent approaches have been devoted to mitigating inconsistencies in feature distribution such as the projection-based aggregation mechanism \cite{pkef}, they do not address the dynamics of the representation space during training (details are illustrated in Appendix \ref{shortcoming_pkef}). \textbf{2) \textit{Differences in label space distribution during backward propagation}}: Existing methods learn information from different behaviors but encounter conflicts during gradient updates due to differences in label distributions. Previous methods \cite{cigf, pkef} have attempted to address gradient conflicts using decoupled inputs, with the issue of gradient coupling in aggregation still existing. These factors can lead to negative transfer problems \cite{negativeTransfer}. 
\end{itemize}

To address these two challenges, we propose a \underline{\textbf{C}}ombinatorial \underline{\textbf{O}}ptimization \underline{\textbf{P}}erspective based \underline{\textbf{F}}ramework for Multi-behavior Recommendation (COPF). It consists of the Combinatorial Optimization Graph Convolution Network (COGCN) and the Distributed Fitting Multi-Expert Networks (DFME). To tackle the combinatorial optimization problem in the fusion step, COGCN restricts the solution space of combinatorial optimization by imposing different degrees of constraints to user behavior patterns across various stages (\textit{Pre-behavior, In-behavior, Post-behavior}) based on graph convolutional networks, thus achieving efficient multi-behavior fusion.

To coordinate the correlations between tasks, DFME improves the forward and backward propagation processes during the multi-behavior prediction phase from two perspectives: feature and label. At the feature level, DFME regards different behaviors as independent tasks, and utilizes contrastive learning to adaptively align the distributions of target and auxiliary behaviors. Considering that behavior aggregation can be affected by differences in behavior feature distributions, DFME incorporates a specialized behavior-fitting expert to refine the representation space for each behavior before aggregation, thereby diminishing distribution bias while maintaining spatial generalization. At the label level, DFME further decouples the gradient between the target and auxiliary behaviors during aggregation, avoiding the influence of other tasks on the gradient updates of the target task. The above designs enable the effective utilization of auxiliary tasks to adjust the model-fitted data distribution to be more consistent with the target task’s test distribution, alleviating the negative transfer problem caused by uncoordinated task relationships (explained in Appendix \ref{method_dfme}).

In summary, the main contributions of our works are as follows:
\begin{itemize}
    \item To our knowledge, we are the first to propose examining the multi-behavior fusion problem from a combinatorial optimization perspective. Specifically, we highlight the benefits of behavioral constraints for multi-behavior recommendation and analyze the limitations of existing methods in this perspective, and then propose Combinatorial Optimization Graph Convolutional Network (COGCN) as a solution. It applies different degrees of constraints to user behavior patterns across various stages, effectively facilitating the process of multi-behavior fusion.
    \item We investigate the limitations of multi-task methods in multi-behavior recommendation from structural perspectives (i.e., the feature and label perspectives) and propose Distributed Fitting Multi-Expert Networks (DFME). It is designed to coordinate task correlations by improving the forward and backward propagation processes, thus alleviating the negative transfer problem in MTL.
    \item We conduct comprehensive experiments on three real-world datasets, demonstrating that our proposed COPF has superior performance in multi-behavior recommendation. Further experimental results also verify the rationality and effectiveness of COGCN and DFME.
\end{itemize}

\section{RELATED WORK}
\label{related_work}
\textbf{Multi-behavior Recommendation.}
Multi-behavior recommendation methods aim to leverage the multiple behavior signals of users to alleviate the issue of data sparsity in target behavior. Early multi-behavior methods extend matrix factorization to accommodate multi-behavior data \cite{mf1,mf2,mf3}, such as CMF \cite{cmf}. Besides, some other methods use auxiliary behavioral signals to design new sampling strategies \cite{bprh,mc-bpr,bpr_resolve, vals}. But none of them exploits deep behavioral information.

With the rise of deep learning \cite{chengqing2023multi,liang2023knowledge,li2024dual,dang2023uniform}, deep neural networks (DNNs) and graph convolutional networks (GCNs) have been proven to be more suitable approaches for multi-behavior fusion, as they can delve deeper into the multiple information of different behaviors. 
DNN-based methods often heavily utilize neural networks to extract information from user-item interactions, and the information is embedded within the representations. For example, DIPN \cite{dipn} and MATN \cite{matn} capture the implicit relationship between behaviors through attention mechanism. NMTR \cite{nmtr} treats all behaviors as prediction targets and transfers prediction results between behaviors. However, most DNN-based models fail to capture high-order relationships, resulting in poor performance. 

In contrast, GCN-based models can learn higher-order relations between users and items, making them the most popular methods for multi-behavior recommendation currently. This type of method mainly learns behavioral information through graph convolution and considers capturing user behavior patterns by holistic modeling of multiple relationships between users and items. S-MBRec \cite{smbrec} models GCN for each behavior individually, focusing solely on the information of the current behavior without learning user patterns. MBGCN \cite{mbgcn}, CML \cite{CML} and CIGF \cite{cigf} further aggregate representations between behaviors through learnable parameters, with no constraints on user behavior patterns. Recent methods CRGCN \cite{crgcn} and MB-CGCN \cite{mbcgcn} take into account the hierarchical correlation between behaviors and employ cascading behavior network, and PKEF \cite{pkef} further considers the bias in cascading networks, but this cascading paradigm imposes overly strict constraints on user behavior patterns. BCIPM \cite{bipn} successively learns global behavior information and target behavior information, which only considers patterns before the target behavior. Therefore, existing methods do not adequately model user behavior patterns.

\textbf{MTL for Recommendation.}
Recent research works have extensively applied multi-task learning (MTL) \cite{stem,li2020improving,moe} methods to recommendation systems to leverage heterogeneous user information. The traditional multi-task learning method is the shared bottom \cite{sharebottom} structure, which shares the bottom network to learn representations and uses separate tower network to predict each task. While some multi-behavior approaches based on this structure \cite{ghcf,crgcn,mbgmn,CML} can achieve knowledge sharing between tasks, their effectiveness may be affected by task differences. To better jointly optimize the model, some methods have introduced attention-based gating mechanisms into MTL structures. For example, MOE \cite{moe} and MMOE \cite{mmoe} propose a multi-expert structure shared by all tasks, and utilize gating networks to obtain expert fusion weights for each task. PLE \cite{PLE} further divides experts into shared experts and task-specific experts. However, the above methods all share inputs between tasks, which can lead to negative information transfer due to gradient conflict. MESI in CIGF \cite{cigf} specifically decouples inputs between tasks, mitigating the negative impact of coupled gradients. PME in PKEF \cite{pkef} further introduces a projection mechanism during task fusion to eliminate harmful information. Nevertheless, none of these approaches fully alleviate the negative transfer problem caused by the aggregation of tasks with differing feature and label distributions from a structural perspective.


\section{PRELIMINARY}
\label{Definition}
In our framework, we use $u$ and $v$ to represent a user and an item, respectively. The user set and item set are denoted as $\mathbf{U}=\{u_1, u_2, \cdots, u_M\}$ and $\mathbf{V}=\{v_1, v_2, \cdots, v_N\}$, where $M$ and $N$ represent the total number of users and items, respectively. The behavior types $k \in \{1,2,...,K\}$ maintain a consistent order among behaviors (i.e., 1 and $K$ correspond to the most upstream and downstream behaviors, respectively). The user-item interaction matrices for the $K$ behavior types can be represented as $\mathcal{B} =\left\{\mathbf{B}_{1},\mathbf{B}_{2},\cdots,\mathbf{B}_{K}\right\}$, where $\mathbf{B}_{k}=\left[b_{(k)uv}\right]_{|\mathbf{U}|\times|\mathbf{V}|}\in \left\{0, 1\right\}$ indicates whether the user $u$ interacts with the item $v$ under behavior $k$. Additionally, we define a user-item bipartite graph $\mathcal{G}=(\mathcal{H}, \mathcal{E}, \mathcal{B})$ to represent the various interaction data between users and items, where $\mathcal{H} = \mathbf{U}\cup\mathbf{V}$, $\mathcal{E} = \cup_{k = 1}^{K}\mathcal{E}_{k}$ is the edge set including all interactions. For multi-behavior recommendation, there is a specific target behavior (e.g., buy) to be optimized, and other behaviors are considered auxiliary behaviors to assist in predicting the target behavior. The target behavior is the most downstream behavior (behavior $K$).

\section{METHODOLOGY}
\begin{figure*}[t]
	\centering
	\setlength{\belowcaptionskip}{-0.0cm}
	\setlength{\abovecaptionskip}{-0.0cm}
	\includegraphics[width=1.0\textwidth]{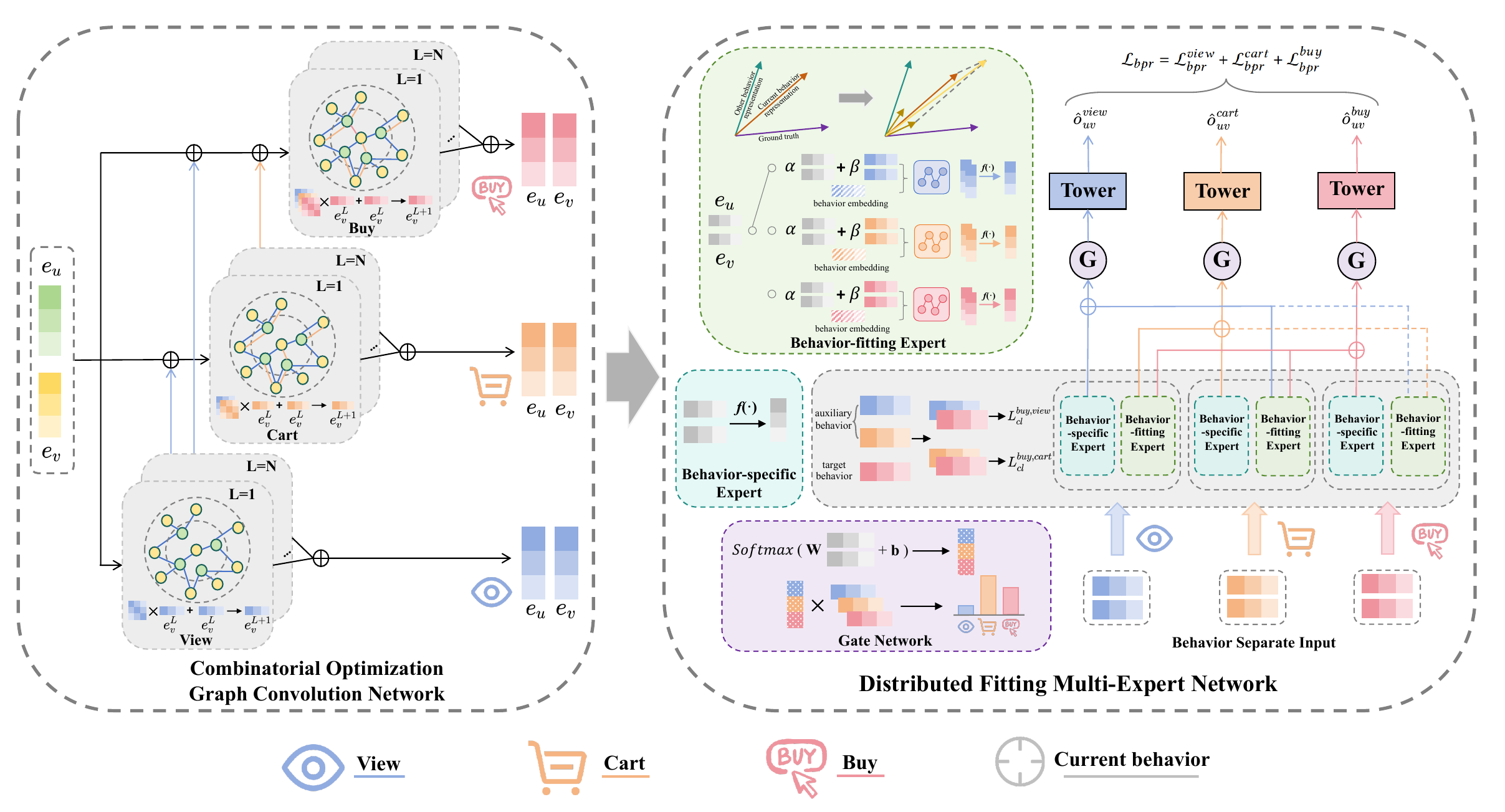}
	\caption{Illustration of the proposed COPF framework, and we use three user behavior types as examples: view, cart, and buy. ($\oplus$) denotes the element-wise addition operation, $f(\cdot)$ represents the function to generate experts, and dotted lines represent stop gradient.}
	\label{fig:framework}
	\vspace{-4mm}
\end{figure*}

We devise a "Combinatorial Optimization Perspective based Framework" (COPF) for multi-behavior recommendation, which contains two parts: (1) Combinatorial Optimization Graph Convolution Network(COGCN); (2) Distributed Fitting Multi-Expert Network(DFME). Figure \ref{fig:framework} illustrates the overall architecture of the proposed framework.

\subsection{Embedding Layer}
We first look up the low-dimensional dense embeddings for user $u$ and item $v$ from the embedding tables using their one-hot vectors, respectively. Specifically, the process for obtaining these embeddings can be formulated as follows:
\begin{equation}
\begin{aligned}
\mathbf{e}_{u}=\mathbf{E}_{u}^{T}\cdot\mathbf{ID}_{u} \in \mathbb{R}^{d}, \mathbf{e}_{v}=\mathbf{E}_{v}^{T}\cdot\mathbf{ID}_{v} \in \mathbb{R}^{d}
\end{aligned}
\vspace{-0.1mm}
\end{equation}
where $\mathbf{ID}_{u}$ and $\mathbf{ID}_{v}$ denotes the one-hot vectors of user $u$ and item $v$, $\mathbf{E}_u \in \mathbb{R}^{|\mathbf{U}| \times d}$ and $\mathbf{E}_v \in \mathbb{R}^{|\mathbf{V}| \times d}$ are the embedding tables for users and items, $|\mathbf{U}|$ and $|\mathbf{V}|$ are the total number of users and items, and $d$ is the embedding size.

\subsection{Combinatorial Optimization Graph Convolution Network}
The recent multi-behavior methods have aimed to deeply explore user behavior patterns to enhance model performance. However, they impose constraints on user behavior patterns that are either too strict or too slack during the fusion step, making it difficult for the model to accurately capture these patterns. 

To solve the above problem, we examine the multi-behavior fusion process from a combinatorial optimization perspective and propose the Combinatorial Optimization Graph Convolutional Network (COGCN). It imposes different degrees of constraints on behavior patterns at different stages of user behavior, so as to learn the optimal behavioral information.

\subsubsection{Definition of Constraints in Combinatorial Optimization.}
To better describe the constraints at different stages of behavior, we first introduce the following definitions:

\textbf{Definition 1} (\textit{User Behavior Pattern}). User behavior patterns are defined as high-frequency behavior chains between users and all items on the platform, representing the user's personalized habits. Formally, for any user $u \in \mathbf{U}$, his/her user behavior pattern is: $u$→$\cdots$→$b_k$→$\cdots$→$\mathbf{V}$, where $b_k$ is the behavior $k$ and $K$ is the number of behavior types. For example, for a user who directly buys items that they find appealing while viewing, his/her user behavior pattern is: $u$→\textit{view}→\textit{buy}→$\mathbf{V}$.


\textbf{Definition 2} (\textit{Upstream and Downstream Behavior}). 
The order of behavior definition is often derived from the mainstream behavior habits of most users in the real world. For example, consider the combination \textit{view → cart → buy}. Upstream behavior refers to the behavior that precedes the current behavior. Similarly, downstream behavior refers to the behavior that follows the current behavior. In the example, \textit{view} is the upstream behavior of \textit{cart}, and \textit{buy} is the downstream behavior of \textit{cart}.



It can be seen that for the number of behavior types $K$, the total number of possible user behavior patterns is $\sum_{i=1}^K\frac{K!}{(K-i)!}$. To restrict the solution space in the combinatorial optimization of multi-behavior recommendation systems, we impose constraints on user behavior patterns at three stages: pre-behavior, in-behavior, and post-behavior. The details are as follows.

\textbf{Definition 3} (\textit{Pre-behavior Constraint}). Many users interact with items in a predetermined order of behaviors. Therefore, the upstream behavior information of the current behavior is essential. Formally, for any behavior $k (k>1)$, 
The input of its encoder $\mathbf{s}_{input}^{k}$ is denoted as 
$\mathbf{s}_{input}^{k}=f(g(\sum_{k^{\prime}=1}^{k - 1}\mathbf{s}_{output}^{k^{\prime}},\mathbf{s}_{init}))$, where 
$\mathbf{s}_{init}$ is the initial input, and $\mathbf{s}_{output}^{k^{\prime}}$ is the information of the behavior $k^{\prime}$, which is actually the output of its encoder. $g(\cdot)$ is the aggregation function (e.g., summation) and $f(\cdot)$ is the transition function (e.g., matrix multiplication, graph convolution). In this way, we constrain the interaction between behaviors.


\textbf{Definition 4} (\textit{In-behavior Constraint}). In-behavior Constraints essentially pertain to the modeling of the current behavior (i.e., the 
transition function $f(\cdot)$). In order to capture more complex or implicit user behavior patterns (for example, a user alternates between \textit{view} and \textit{cart} before \textit{buy}), we similarly utilize GCN with heterogeneous relations \cite{mbgcn}.  Meanwhile, to prevent poor model generalization performance and overfitting problems caused by information leakage, we define that the current behavior node learning process cannot contain semantic information of downstream behaviors. Specifically, for the current behavior $k$, we have the output of its encoder:
\begin{equation}
\mathbf{s}_{output}^{k}=Agg(\mathbf{s}_{input}^{k},\{\mathbf{B}_{k^{\prime}}|\mathbf{B}_{k^{\prime}}\in \mathcal{B},1\leq k^{\prime}\leq k\})
\end{equation}
where $\mathbf{s}_{input}^{k} = g(\sum_{k^{\prime}=1}^{k - 1}\mathbf{s}_{output}^{k^{\prime}},\mathbf{s}_{init})$.

\textbf{Definition 5} (\textit{Post-behavior Constraint}). For each behavior, we design decoupled outputs to partition the solution space, modeling user behavior patterns that end with different behaviors. Through joint optimization, the model can achieve better performance. This is consistent with the perspective of our proposed DFME.

\subsubsection{Graph Convolution.}
\label{para_inter_enhance}
In the previous part, we outlined the specific constraints for different behavioral stages. As graph convolutional networks (GCNs) can efficiently utilize high-order connectivity between users and items, we use a GCN-based paradigm to model the multi-behavior information fusion in the combinatorial optimization perspective.

Given the adjacency matrices of different behaviors, we modify them to meet the requirements of graph convolution:
\begin{equation}
\mathbf{A}_{k}=\left(\begin{array}{cc}
0 & \mathbf{B}_{k} \\
\left(\mathbf{B}_{k}\right)^{T} & 0
\end{array}\right)
\end{equation}
where $\mathbf{A}_{k}$ is the adjacency matrix of behavior $k$ in the graph. For the same purpose, we obtain the embedding matrices for users and items, respectively:
\begin{equation}
\mathbf{E}_{u}=[\begin{array}{cc}\mathbf{e}_{u_{1}},\cdots,\mathbf{e}_{u_{|\mathrm{U}|}}\end{array}], 
\mathbf{E}_{v}=[\begin{array}{cc}\mathbf{e}_{v_{1}},\cdots,\mathbf{e}_{v_{|\mathrm{V}|}}\end{array}],
\end{equation}
we then capture the interaction information of behaviors through graph convolution. Inspired by \cite{lightgcn,mbgcn}, for behavior $k$, we have:
\begin{equation}
\label{mess_gcn}
\mathbf{E}^{k,l+1}=\sum_{k^{\prime}=1}^{k}(\mathbf{D}^{-1}\mathbf{A}_{k^{\prime}} + \mathbf{I})\mathbf{E}^{k,l}
\end{equation}
where $\mathbf{D}$ is the diagonal identity matrix, $\mathbf{I}$ denotes an identity matrix. $\mathbf{E}^{k,l} = \mathbf{E}_{u}^{k,l}||\mathbf{E}_{v}^{k,l}$, $(||)$ is the concatenate operation and $l$ denotes the $l$-th layer. the initial input of the model $\mathbf{E}^{1,0} = \mathbf{E}_{u}||\mathbf{E}_{v}$. We utilize the adjacency matrices of the current behavior and its upstream behavior for message propagation and aggregation on each layer $l$, thereby implementing the \textit{In-behavior Constraint}.

Further, in order to implement the \textit{Pre-behavior Constraint}, we define the hierarchical information transfer between behaviors as:
\begin{equation}
\mathbf{E}^{k+1,0}=\sum_{k^{\prime}=1}^{k}\mathbf{E}^{k^{\prime},L}+\mathbf{E}^{1,0}
\end{equation}
where $L$ denotes the total layers of GCN. Here, we combine the last layer representations of each upstream behavior representation with the initial representation as the input of the current behavior.

Follow the \textit{Post-behavior Constraint}, We independently output the representations of each behavior for subsequent multi-task learning. To be specific, we directly add the outputs of different layers to get relations of different orders. For the embeddings of user $u$ and item $v$ in 
$\mathbf{E}^{k,l}$($\mathbf{e}_u^{k,l}$ and $\mathbf{e}_v^{k,l}$), we have:
\begin{equation}
\mathbf{e}_u^{k,*} = \sum_{l=0}^{L}\mathbf{e}_u^{k,l}, \mathbf{e}_v^{k,*} = \sum_{l=0}^{L}\mathbf{e}_v^{k,l}
\end{equation}
where $L$ is the number of GCN layers. 
\subsection{Distributed Fitting Multi-Expert Network}
By employing COGCN in the multi-behavior fusion step, we have obtained representations for user $u$ and item $v$ under each behavior $k$. The subsequent task is to devise a proper structure for multi-behavior prediction. Many methods \cite{pkef,cigf,crgcn} have utilized MTL modules to fully leverage multi-behavior information to assist in predicting target behavior, which has demonstrated their effectiveness. However, these MTL methods exhibit insufficient exploration in their structural design, failing to account for the potential negative transfer effects caused by differences in feature and label distributions during the learning process. To handle the drawbacks of the existing MTL modules, we propose the Distributed Fitting Multi-Expert Network(DFME), which controls behavior interactions through both features and labels, thereby coordinating the relationships between tasks. The specific details are as follows.

\subsubsection{Generating of Behavior-specific Experts.}
As contrastive learning can alleviate distributional biases between different data sources, we utilize it to adaptively learn the distributional similarity between target behaviors and auxiliary behaviors before generating experts. Take the auxiliary behavior $k$ as an example, we have:
\begin{equation}
\mathcal{L}_{cl,U}^{K,k}=\frac{1}{|\mathbf{U}|}\sum_{u\in\mathbf{U}}-log\frac{\exp(\varphi(\mathbf{e}_u^{K,*},\mathbf{e}_u^{k,*})/\tau)}{\sum_{u^{\prime}\in\mathbf{U}}\exp(\varphi(\mathbf{e}_u^{K,*},\mathbf{e}_{u^{\prime}}^{k,*})/\tau)}
\end{equation}
where $\tau$ represents the temperature hyperparameter for the softmax function, and 
$\varphi(\cdot)$ is a function for calculating the similarity between two vectors(e.g., inner product) .The item side follows the same contrastive learning process. Thus, for behavior $k$, the final contrastive loss is $\mathcal{L}_{cl}^{K,k} =\mathcal{L}_{cl,U}^{K,k}+\mathcal{L}_{cl,V}^{K,k}$.

Then, we follow previous methods\cite{cigf} by using decoupled behavior representations to generate behavior-specific experts, thereby preventing gradient conflicts caused by coupled inputs:
\begin{equation}
\mathbf{e}^{k} = \mathbf{e}_u^{k,*} \circ \mathbf{e}_v^{k,*}
\end{equation}
where $(\circ)$ is the hadamard product operation. Since decoupled behavior representations are  utilized to generate experts, we can obtain a total of $k$ behavior-specific experts.

\subsubsection{Generating of Behavior-fitting Experts.}
To address the challenge of mitigating feature distribution bias, we also define a dedicated behavior-fitting expert for each task, whose outputs are used in the subsequent aggregation process. Specifically, for the current behavior $k$ and any other behavior $k^{\prime}$, it can be formulated as:
\begin{equation}
\left\{\begin{array}{c}
\begin{aligned}
\mathbf{e}_{u,in}^{k, k^{\prime}}&=(\alpha \mathbf{e}_u^{k,*}+\beta \mathbf{e}_u^{k^{\prime},*})/2\\
\mathbf{e}_{v,in}^{k, k^{\prime}}&=(\alpha \mathbf{e}_v^{k,*}+\beta \mathbf{e}_v^{k^{\prime},*})/2\\
\mathbf{e}_{in}^{k, k^{\prime}} &= \mathbf{e}_{u,in}^{k, k^{\prime}}||\mathbf{e}_{v,in}^{k, k^{\prime}}\\
\mathbf{e}_{out}^{k, k^{\prime}} &= Agg(\mathbf{e}_{in}^{k, k^{\prime}}, \mathbf{A}_{k^{\prime}})
\end{aligned}
\end{array}\right.
\end{equation}
where $\mathbf{e}_{out}^{k, k^{\prime}} = \mathbf{e}_{u,out}^{k, k^{\prime}}||\mathbf{e}_{v,out}^{k, k^{\prime}}$, $(||)$ is the concatenate operation, $\mathbf{e}_u^{k,*}$ and $\mathbf{e}_u^{k^{\prime},*}$ are the user representations of the $k$- and $k^{\prime}$-th behavior($k^{\prime} \neq k$) respectively, similarly for items. $\alpha, \beta$ are coefficients that control the scaling of behavioral representations, and $Agg(\cdot)$ is a graph convolution operator. In particular, the values of $\alpha$ and $\beta$ should be small to achieve the effect of fine-tuning the representation space. We use a graph convolutional network with the $k^{\prime}$-th behavior interaction matrix (i.e., $Agg(\cdot)$) to capture the effective information contained within the representation $\mathbf{e}_{in}^{k, k^{\prime}}$. Similar to Equation \ref{mess_gcn}, the graph convolutional operator is defined as follows:
\begin{equation}
\mathbf{E}^{k, k^{\prime}, l+1}=(\mathbf{D}^{-1}\mathbf{A}_{k^{\prime}} + \mathbf{I})(\mathbf{E}^{k,k^{\prime}, l}\circ\mathbf{R}^{k^{\prime},l}),\mathbf{R}^{k^{\prime},l}=\mathbf{W}^{l}\mathbf{R}^{k^{\prime},l - 1}
\end{equation}
where ($\circ$) is the hadamard product operation, $\mathbf{A}_{k^{\prime}}$ is the adjacency matrix of behavior $k^{\prime}$. $\mathbf{W}^{l}$ is the layer specific parameter shared with the layer of GCN. We use a hierarchically updated behavior embedding matrix $\mathbf{R}^{k^{\prime},l}$ for fully learning of the $k^{\prime}$-th behavior information, with its initial state is $\mathbf{R}^{k^{\prime},1}$.

As $\mathbf{e}_{out}^{k, k^{\prime}}$ contains the representation of users and items, we utilize the hadamard product operation to generate final experts, which is similar to behavior-specific experts:
\begin{equation}
\mathbf{e}^{k, k^{\prime}} = \mathbf{e}_{u,out}^{k, k^{\prime}} \circ \mathbf{e}_{v,out}^{k, k^{\prime}}
\end{equation}

\subsubsection{Aggregating of Experts.}
In order to mitigate the problem of negative transfer caused by distribution differences between features and labels, we improve the task aggregation mechanism in both forward and backward propagation. As we can see, behavior-aware graph convolution operation and representation scaling mechanism help us capture the effective components of other behaviors, which can be further utilized to alleviate the negative transfer caused by feature distribution differences in gated aggregation. To be specific, We define the gate for task $k$ as:
\begin{equation}
\mathbf{g}^k=Softmax(\mathbf{W}_g(\mathbf{e}_{u}^{k,*}||\mathbf{e}_{v}^{k,*})+\mathbf{b}_g)
\end{equation}
where $(||)$ is the concatenate operation, $\mathbf{W}_{g} \in \mathbb{R}^{K \times 2d}$ and $\mathbf{b}_{g} \in \mathbb{R}^{K \times 1}$ are feature transformation matrix and bias matrix, and $\mathbf{g}^{k} \in \mathbb{R}^{K \times 1}$ is the attention vector which are used as selector to calculate the weighted sum of all experts. We then take the refined representations $\mathbf{e}^{j, k} ( j \in \{1,2,...,K\} \cap j \neq k)$ of other behaviors and $\mathbf{e}^{k}$ as targets of aggregation by the $k$-th gate. In order to eliminate the negative impact caused by the distribution differences in labels between the auxiliary and target behaviours, we ensure that the parameters of the target task are not updated by the gradient updates from the auxiliary tasks, thus preventing interference from auxiliary behaviors. Formally, we have:
\begin{equation}
\mathbf{o}^{k}(j)=\left\{\begin{array}{c}
\begin{aligned}
& \mathbf{g}^{k}(j) \cdot \mathbf{e}^{k},& j& = k\\
& \mathbf{g}^{k}(j) \cdot \mathbf{e}^{j, k},& j&\neq k\\
& sg(\mathbf{g}^{k}(j) \cdot \mathbf{e}^{j, k}),& j&\neq k \,and\, j = K
\end{aligned}
\end{array}\right.
\end{equation}
where $\mathbf{g}^{k}(j)$ denotes the $j$-th element of vector $\mathbf{g}^{k}$. $sg(\cdot)$ is the stop gradient operation. The final prediction for task $k$ is calculated as:
\begin{equation}
\hat{o}_{uv}^{k} = h^{k}(\sum_{j=1}^{K} {\mathbf{o}^{k}(j)})
\end{equation}
where $h^{k}(\cdot)$ is the tower function. For simplicity, we use average operation as the tower function here. $\hat{o}_{uv}^{k}$ is the prediction score of whether user $u$ will have interaction with item $v$ under behavior $k$.

\subsection{Joint Optimization}
As we have obtained the final prediction $\hat{o}_{uv}^{k}$ for each behavior 
$k$, we leverage the Bayesian Personalized Ranking (BPR)\cite{bpr} loss to optimize the model:
\begin{equation}
\mathcal{L}_{bpr} = -\sum_{k=1}^{K}\sum_{(u,s,t)\in \mathcal{O}_k} \lambda_k * \textup{ln} \sigma(\hat{o}_{us}^{k} - \hat{o}_{ut}^{k})
\end{equation}
where $\mathcal{O}_k = \left\{(u,s,t)|(u,s)\in \mathcal{O}_k^{+}, (u,t) \in \mathcal{O}_k^{-} \right\}$ denotes the training dataset. $\mathcal{O}_k^+$ and $\mathcal{O}_k^-$ indicates the observed and unobserved user-item interactions under behavior $k$, respectively. $\lambda_k$ is the coefficient of behavior $k$, and $\sigma$ is the sigmoid function.

In all, the final loss can be formulated as:
\begin{equation}
    \mathcal{L}(\Theta) = \mathcal{L}_{bpr} + \gamma \sum_{k=1}^{K - 1}\mathcal{L}_{cl}^{K,k} + \mu ||\Theta||^2_2
\end{equation}
where $\gamma$ is the coefficient of cl loss, $\Theta$ represents set of all model parameters. $\mu$ is the $L_2$ regularization coefficient for $\Theta$.


\section{EXPERIMENTS}
\label{experiments}

\begin{table}[t]
\setlength{\abovecaptionskip}{0cm}
\setlength{\belowcaptionskip}{0mm}
\caption{Statistics of evaluation datasets.}
\centering
\resizebox{\linewidth}{!}{
\begin{tabular}{c|ccccc}
\toprule
Dataset & \#User & \#Item & \#Interaction & \#Target Interaction &  \#Interactive Behavior Type \\ \midrule
Beibei & 21,716 & 7,977 & $3.3 \times 10^6$ & 282,860 & \{View,Cart,Buy\} \\ 
Taobao & 15,449 & 11,953 & $1.2 \times 10^6$ & 92,180 & \{View,Cart,Buy\} \\
Tmall & 41,738 & 11,953 & $2.3 \times 10^6$ & 255,586 & \{View,Collect,Cart,Buy\} \\  \bottomrule

\end{tabular}
}
\vspace{-3mm}
\label{dataset}
\end{table}

\subsection{Experimental Setting}
\subsubsection{Parameter Settings}
Our proposed COPF is implemented in TensorFlow \cite{TensorFlow}. For a fair comparison, following PKEF \cite{pkef} and BCIPM \cite{bipn}, we set the embedding size to 64. We initialize the parameters using Xavier \cite{xavier}. The parameters are optimized by Adam \cite{adam}, while the learning rate is set to $10^{-3}$. We search the number of GCN layers in \{1,2,3,4\}. Moreover, we adjust the loss coefficients for each behavior in \{0,1/6,2/6,3/6,4/6,5/6,1\} and fix the sum of the coefficients for all actions as 1. The coefficient of contrastive loss $\gamma$ and $L_2$ regularization $\mu$ are set to 1 and 0.01, respectively. All experiments are run 5 times and average results are reported. For fairness, the parameter settings of the baseline are adjusted and searched by referring to the original work. Furthermore, we conduct parameter analysis experiments, which are shown in Section \ref{hyper}.

\subsubsection{Dataset Description}
We use three public datasets (Beibei, Taobao and Tmall) to validate the effectiveness of our proposed COPF model. The pre-processing of these datasets is consistent with the previous methods \cite{cigf,pkef}. Specifically, we eliminate duplicate user-item interactions through retaining the earliest one. The statistical information of these three datasets is summarized in Table \ref{dataset}. 

\subsubsection{Evaluation Metrics}
To evaluate the performance of COPF and baseline methods in top-k item recommendation, we use two metrics: Hit Ratio (\textit{HR@K}) and Normalized Discounted Cumulative Gain (\textit{NDCG@K}). In all our experiments, we set $K = 10$.

\subsubsection{Baseline Models}
To validate the effectiveness of COPF, we compared it with numerous baseline models in recent years, which can be divided into three categories: \textbf{(1) Single-behavior methods:} MF-BPR \cite{bpr}, NeuMF \cite{ncf} and LightGCN \cite{lightgcn}, \textbf{(2) Multi-behavior methods without MTL:} RGCN \cite{RGNN}, GNMR \cite{gnmr}, NMTR \cite{nmtr}, \\ MBGCN\footnote{https://github.com/tsinghua-fib-lab/MBGCN} \cite{mbgcn}, S-MBRec \cite{smbrec}, KMCLR \cite{kmclr} and MB-CGCN\footnote{https://github.com/SS-00-SS/MBCGCN} \cite{mbcgcn}, \textbf{(3) Multi-behavior methods with MTL:} CML \cite{CML}, CRGCN \cite{crgcn}, CIGF \cite{cigf}, PKEF \footnote{https://github.com/MC-CV/PKEF} \cite{pkef} and BCIPM\footnote{https://github.com/MingshiYan/BIPN} \cite{bipn}.

\begin{table}[t]
\setlength{\abovecaptionskip}{0cm}
\setlength{\belowcaptionskip}{0mm}
\caption{The overall performance comparison. Boldface denotes the highest score and underline indicates the results of the best baselines. $\star$ represents significance level $p$-value $<0.05$ of comparing COPF with the best baseline.}
    \centering
    \begin{threeparttable}
	\resizebox{\linewidth}{!}{
    \begin{tabular}{c|cccccc}
    \toprule
    \multirow{2}{*}{Model}&
    \multicolumn{2}{c}{Beibei}&\multicolumn{2}{c}{Taobao}&\multicolumn{2}{c}{Tmall}\cr
    \cmidrule(lr){2-3} \cmidrule(lr){4-5} \cmidrule(lr){6-7} 
    &HR&NDCG&HR&NDCG&HR&NDCG\cr
    \midrule
    MF-BPR&0.0191&0.0049&0.0076&0.0036&0.0230&0.0207\cr
    NeuMF&0.0232&0.0135&0.0236&0.0128&0.0124&0.0062\cr
    LightGCN&0.0391&0.0209&0.0411&0.0240&0.0393&0.0209\cr
    \hline
    RGCN&0.0363&0.0188&0.0215&0.0104&0.0316&0.0157\cr
    GNMR&0.0413&0.0221&0.0368&0.0216&0.0393&0.0193\cr
    NMTR&0.0429&0.0198&0.0282&0.0137&0.0536&0.0286\cr
    MBGCN&0.0470&0.0259&0.0509&0.0294&0.0549&0.0285\cr
    S-MBRec&0.0489&0.0253&0.0498&0.0269&0.0694&0.0362\cr
    KMCLR&0.0531&0.0263&0.1185&0.0659&0.0603&0.0310\cr
    MB-CGCN&0.0579&0.0381&0.1233&0.0677&0.0984&0.0558\cr
    \hline
    CML&0.0542&0.0268&0.1203&0.0661&0.0448&0.0227\cr
    CRGCN&0.0459&0.0324&0.0855&0.0439&0.0840&0.0442\cr
    CIGF&0.0809&0.0400&0.0897&0.0474&0.1150&0.0636\cr
    PKEF&\underline{0.1130}&\underline{0.0582}&\underline{0.1385}&\underline{0.0785}&0.1277&0.0721\cr
    BCIPM&0.0458&0.0221&0.1201&0.0656&\underline{0.1414}&\underline{0.0741}\cr
    \hline
    \textbf{COPF}&\textbf{0.1694}$^\star$&\textbf{0.0903}$^\star$&\textbf{0.1552}$^\star$&\textbf{0.0838}$^\star$&\textbf{0.1755}$^\star$&\textbf{0.0967}$^\star$\cr
    \hline
    Rel Impr.&49.91\%&55.15\%&12.06\%&6.75\%&24.12\%&30.50\%\cr
    \bottomrule
    \end{tabular}}
    \end{threeparttable}
    \vspace{-4mm}
    \label{comparisons_model}
\end{table}

\subsection{Performance Comparison}
\label{performance}
Table \ref{comparisons_model} shows the performance of methods on three datasets with respect to HR@10 and NDCG@10. 
We have the following findings:
\begin{itemize}

\item Our proposed COPF model achieves the best performance on all three datasets. Specifically, COPF improves the best baselines by \textbf{49.91$\%$}, \textbf{12.06$\%$}, and \textbf{ 24.12$\%$} in terms of HR ( \textbf{55.15$\%$}, \textbf{6.75$\%$}, and \textbf{30.50$\%$} in terms of NDCG) on Beibei, Taobao, and Tmall datasets, respectively. Due to the varying user behavior patterns across different datasets, the superior performance on all datasets further demonstrates the applicability and effectiveness of COPF for multi-behavior recommendation.

\item In single-behavior methods, LightGCN achieves better performance than MF-BPR and NeuMF, while in multi-behavior methods, MBGCN also outperforms NMTR. This demonstrates the advantage of GCNs in capturing high-order interactive information. Furthermore, most of the multi-behavior recommendation methods, such as MBGCN, perform better than single-behavior methods on all three datasets, which highlights the superiority of leveraging multi-behavior information for learning. Finally, the excellent performance of CML and KMCLR also illustrates the effectiveness of contrastive learning.

\item Although models vary in network structure, the multi-behavior methods with MTL generally perform better overall compared to those without MTL. For example, PKEF consistently outperforms all the multi-behavior methods without MTL. It is worth noting that KMCLR and MB-CGCN perform better among the multi-behavior methods without MTL. The possible reasons are that KMCLR enhances the original multi-behavioral information by introducing external knowledge graph information; Meanwhile, MB-CGCN reduces the solution space of the multi-behavior fusion problem through cascade constraints. Even though this constraint is overly strict, it still achieves relatively better results in the biased space.

\item MBGCN outperforms RGCN by considering the contribution of each behavior during behavior fusion. Compared to them, CIGF utilizes multi-task learning in the prediction process, which further improves the performance. However, they still lacked
proper constraints or imposed overly relaxed constraints on user behavior patterns. Recent approaches like CRGCN, MB-CGCN, and PKEF use cascading paradigm to constrain the learning of user behavior patterns; BCIPM further relaxes the constraints within the cascading paradigm and highlights the significance of the target behavior. As we can see, PKEF achieves second performance only to our model on Beibei and Taobao, while BCIPM does the same on Tmall. This indicates the necessity of considering user behavior pattern constraints from a combinatorial optimization perspective.
\end{itemize}

\subsection{Ablation Study}
\label{ablation_study}
\subsubsection{Impact of the Key Components}

\begin{table}[t]
    \setlength{\abovecaptionskip}{0cm}
    \setlength{\belowcaptionskip}{0mm}
    \caption{Performances of different COPF variants.}
    \centering
    \begin{threeparttable}
    \resizebox{\linewidth}{!}{
    \begin{tabular}{c|cccccc}
    \toprule
    \multirow{2}{*}{Model}&
    \multicolumn{2}{c}{Beibei}&\multicolumn{2}{c}{Taobao}&\multicolumn{2}{c}{Tmall}\cr
    \cmidrule(lr){2-3} \cmidrule(lr){4-5} \cmidrule(lr){6-7}
    &HR&NDCG&HR&NDCG&HR&NDCG\cr
    \midrule
    w/o COGCN &0.0601&0.0294&0.0801&0.0417&0.0783&0.0437\cr
    COPF-P &0.1282&0.0661&0.1041&0.0540&0.1201&0.0639\cr
    COPF-A &0.1649&0.0885&0.1389&0.0740&0.1592&0.0883\cr
    COPF-D &0.0333&0.0163&0.1052&0.0595&0.1064&0.0606\cr
    COPF-F &0.1285&0.0665&0.0996&0.0515&0.1167&0.0620\cr
    COPF-C &0.1514&0.0861&0.0913&0.0511&0.0785&0.0446\cr
    COPF-B &0.1622&0.0870&0.1531&0.0817&0.1714&0.0957\cr
    COPF-H &0.1199&0.0649&0.0841&0.0478&0.0809&0.0475\cr
    \hline
    w/o DFME &0.0821&0.0400&0.1120&0.0607&0.1146&0.0643\cr
    w/o con. &0.1154&0.0581&0.1476&0.0788&0.1425&0.0791\cr
    w/o for. &0.1649&0.0886&0.0471&0.0247&0.1391&0.0769\cr
    w/o back. &0.1676&0.0886&0.1446&0.0781&0.1601&0.0888\cr
    all sg. &0.1656&0.0888&0.1508&0.0809&0.1553&0.0866\cr
    w/o fit. &0.1182&0.0614&0.0827&0.0438&0.1346&0.0738\cr
    \hline
    \textbf{COPF}&\textbf{0.1694}&\textbf{0.0903}&\textbf{0.1552}&\textbf{0.0838}&\textbf{0.1755}&\textbf{0.0967}\cr
    \bottomrule
    \end{tabular}}
    \end{threeparttable}
    \vspace{-4mm}
    \label{tab:ablation_key}
\end{table}
To evaluate the effectiveness of sub-modules in our COPF framework, we conducted ablation experiments on COGCN and DFME respectively. For COGCN, we mainly consider the constraints of each stage. Specifically, we first define two strict constraints: (1) \textit{strict pre-behavior constraint}: Change the pre-behavior constraint to "only receive information about the most recent upstream behavior"; (2) \textit{strict in-behavior constraint}: Change the in-behavior constraint to "only learn the current behavior signal". Then we design the following variants: (1) \textbf{w/o COGCN}: Replace COGCN with multiple LightGCN.(2) \textbf{COPF-P}: Remove pre-behavior constraint.(3) \textbf{COPF-A}: Remove in-behavior constraint.(4) \textbf{COPF-D}: Remove post-behavior constraint(and no DFME either).(5) \textbf{COPF-F}: Remove pre-behavior and in-behavior constraints.(6) \textbf{COPF-C}:Use a strict cascading paradigm.(7) \textbf{COPF-B}: Use strict pre-behavior constraint instead. (8) \textbf{COPF-H}: Use strict in-behavior constraint instead. For DFME, we have: (1) \textbf{w/o DFME}: Replace DFME with the bilinear module.(2) \textbf{w/o con.}: Remove contrastive learning.(3) \textbf{w/o for.}: Remove improvement in forward propagation during aggregation.(4) \textbf{w/o back.}: Remove improvement in backward propagation during aggregation.(5) \textbf{all sg.}: Use simple stop-gradient strategy in backward propagation during aggregation.(6) \textbf{w/o fit.}: Remove improvements in both propagation during aggregation. The results in Table \ref{tab:ablation_key} lead to the following conclusions:
\begin{itemize}
\item Comparing the performance of COPF with other variants that modify constraints in COGCN (for fairness, \textbf{COPF-D} is compared with \textbf{w/o DFME}), we observe that removing or altering the constraints of any stage leads to varying degrees of performance degradation. Additionally, \textbf{w/o COGCN} achieves the worst performance on the three datasets compared to other variants. These demonstrate the effectiveness of addressing multi-behavior fusion from a combinatorial optimization perspective and validate the rationality of the constraints established at each stage of COGCN.
\item The performance of each variant modified for DFME is also affected to varying degrees, with the \textbf{w/o DFME} variant performing the worst. This demonstrates the rationality and effectiveness of our proposed DFME.

\end{itemize}

\subsubsection{Impact of the Aggregation Schemes}
\label{knowledge_fusion}
\begin{table}[t]
    \setlength{\abovecaptionskip}{0cm}
    \setlength{\belowcaptionskip}{0mm}
    \caption{Performances of different aggregation schemes.}
    \centering
    \begin{threeparttable}
    \resizebox{\linewidth}{!}{
    \begin{tabular}{c|cccccc}
    \toprule
    \multirow{2}{*}{Model}&
    \multicolumn{2}{c}{Beibei}&\multicolumn{2}{c}{Taobao}&\multicolumn{2}{c}{Tmall}\cr
    \cmidrule(lr){2-3} \cmidrule(lr){4-5} \cmidrule(lr){6-7}
    &HR&NDCG&HR&NDCG&HR&NDCG\cr
    \midrule
    No Aggregation &0.1103 & 0.0560 & 0.1143 & 0.0634 & 0.1165 & 0.0655 \cr
    Summation &0.1163 & 0.0605 & 0.0789 & 0.0410 & 0.1323 & 0.0732\cr
    Linear Trans.&0.0724 & 0.0342 & 0.1317 & 0.0712 & 0.1408 & 0.0771\cr
    Vanilla Fusion &0.1228 & 0.0637 & 0.1080 & 0.0589 & 0.1337 & 0.0749\cr
    Projection Fusion&0.1299 & 0.0672 & 0.1153 & 0.0611 & 0.1442 & 0.0791\cr
    \hline
    \textbf{COPF}&\textbf{0.1694}&\textbf{0.0903}&\textbf{0.1552}&\textbf{0.0838}&\textbf{0.1755}&\textbf{0.0967}\cr
    \bottomrule
    \end{tabular}}
    \end{threeparttable}
    \vspace{-4mm}
    \label{tab:ablation_fusion}
\end{table}

To further explore the optimal approach for coordinating tasks, we compare the proposed scheme with several other alternatives: (1) \textbf{No Aggregation}: No aggregation process between tasks. (2) \textbf{Summation}: Simply add the representation of different tasks. (3) \textbf{Linear Trans.}: Apply a linear transformation to transfer the task representation. (4)\textbf{Vanilla Fusion} \cite{pkef}: Utilize a variant of vanilla attention\cite{atrank}. (5) \textbf{Projection Fusion} \cite{dumn}: Explicitly extract the information through projection mechanism. It is worth noting that in order to directly compare the performance of the aggregation methods, all schemes have incorporated the contrastive learning between behaviors. As shown in Table \ref{tab:ablation_fusion}, We can observe that No Aggregation performs the worst overall among all the schemes, which fully demonstrates the importance of aggregation between tasks. Summation negatively impacts the distribution of representations, leading to relatively poor performance on the three datasets. Additionally, both Linear Trans. and Vanilla Fusion methods ignore the noise that may be introduced during aggregation, which can result in negative information transfer. Projection Fusion utilizes a projection mechanism, avoiding the introduction of harmful information while also mitigating the impact of distribution differences in representations. Finally, our proposed method demonstrates the best performance across all three datasets, highlighting the effectiveness of our scheme.

\subsubsection{Impact of the MTL module}
To further demonstrate the superiority of our proposed DFME in MTL, we compare it with some other MTL models: Shared Bottom \cite{sharebottom}, Bilinear \cite{ghcf}, MMOE \cite{mmoe}, PLE \cite{PLE}, MESI \cite{cigf} and PME \cite{pkef}. These MTL models are applied on top of COGCN for multi-behavior recommendation, which are named COGCN+SB, COGCN+MMOE, COGCN+PLE, COGCN+Bilinear, COGCN+MESI, and COGCN+PME respectively. In particular, We weight the $K$ separate representations generated by COGCN to meet the shared input requirement of the classical MTL models (i.e., Shared Bottom, MMOE, and PLE). The experimental results are presented in Table \ref{tab:ablation_mtl}. COGCN+SB achieves the poorest performance among all MTL models on all datasets. COGCN+MMOE and COGCN+PLE outperform COGCN+SB under the same conditions through gating mechanism, which demonstrates the necessity of task aggregation. COGCN+Bilinear shows better performance by using decoupled input and streamlined task prediction tower functions. COGCN+MESI further combines both decoupled inputs and task aggregation but performs worse than COGCN+Bilinear on Tmall dataset, which is possibly due to differences in feature distributions during aggregation. COGCN+PME further enforces alignment of task representation spaces during aggregation, significantly enhancing performance. Finally, our DFME consistently outperforms all other models on all datasets, verifying its effectiveness for MTL.

\begin{table}[t]
    \setlength{\abovecaptionskip}{0cm}
    \setlength{\belowcaptionskip}{0mm}
    \caption{Performances of different MTL module.}
    \centering
    \begin{threeparttable}
    \resizebox{\linewidth}{!}{
    \begin{tabular}{c|cccccc}
    \toprule
    \multirow{2}{*}{Model}&
    \multicolumn{2}{c}{Beibei}&\multicolumn{2}{c}{Taobao}&\multicolumn{2}{c}{Tmall}\cr
    \cmidrule(lr){2-3} \cmidrule(lr){4-5} \cmidrule(lr){6-7}
    &HR&NDCG&HR&NDCG&HR&NDCG\cr
    \midrule
    COGCN+SB & 0.0328 & 0.0159 & 0.0594 & 0.0305 & 0.0740 & 0.0403\cr
    COGCN+MMOE & 0.0557 & 0.0280 & 0.0610 & 0.0317 & 0.0838 & 0.0440\cr
    COGCN+PLE & 0.0546 & 0.0269 & 0.0649 & 0.0338 & 0.0801 & 0.0422\cr
    COGCN+Bilinear & 0.0629 & 0.0306 & 0.0986 & 0.0522 & 0.1559 & 0.0858\cr
    COGCN+MESI & 0.0885 & 0.0439 & 0.1039 & 0.0540 & 0.1487 & 0.0811\cr
    COGCN+PME & 0.1137 & 0.0572 & 0.1525 & 0.0800 & 0.1572 & 0.0862\cr
    \hline
    \textbf{COGCN+DFME}&\textbf{0.1694}&\textbf{0.0903}&\textbf{0.1552}&\textbf{0.0838}&\textbf{0.1755}&\textbf{0.0967}\cr
    \bottomrule
    \end{tabular}}
    \end{threeparttable}
    \vspace{-4mm}
    \label{tab:ablation_mtl}
\end{table}

\subsection{Compatibility Analysis}

\begin{table}[t]
    \setlength{\abovecaptionskip}{0cm}
    \setlength{\belowcaptionskip}{0mm}
    \caption{Compatibility performance of DFME with different models as backbones ("X+DFME" means using X to replace the COGCN in COPF).}
    \centering
    \begin{threeparttable}
    \resizebox{\linewidth}{!}{
    \begin{tabular}{c|cccccc}
    \toprule
    \multirow{2}{*}{Model}&
    \multicolumn{2}{c}{Beibei}&\multicolumn{2}{c}{Taobao}&\multicolumn{2}{c}{Tmall}\cr
    \cmidrule(lr){2-3} \cmidrule(lr){4-5} \cmidrule(lr){6-7}
    &HR&NDCG&HR&NDCG&HR&NDCG\cr
    \midrule
    LightGCN$_M$&0.0456&0.0224&0.0452&0.0246&0.0489&0.0297\cr
    LightGCN$_M$+DFME & \textbf{0.0928} & \textbf{0.0461} & \textbf{0.0928} & \textbf{0.0522} & \textbf{0.0865} & \textbf{0.0493}\cr
    \hline
    MB-CGCN&0.0579&0.0381&0.1233&0.0677&0.0984&0.0558\cr
    MB-CGCN+DFME & \textbf{0.1356} & \textbf{0.0747} & \textbf{0.1400} & \textbf{0.0764} & \textbf{0.1094} & \textbf{0.0585}\cr
    \Xhline{1px}
    CRGCN&0.0459&0.0324&0.0855&0.0439&0.0840&0.0442\cr
    CRGCN+DFME & \textbf{0.1253} & \textbf{0.0663} & \textbf{0.1322} & \textbf{0.0698} & \textbf{0.1308} & \textbf{0.0736}\cr
    \hline
    CIGF&0.0809&0.0400&0.0897&0.0474&0.1150&0.0636\cr
    CIGF+DFME & \textbf{0.0851} & \textbf{0.0435} & \textbf{0.1097} & \textbf{0.0600} & \textbf{0.1219} & \textbf{0.0704}\cr
    \hline
    PKEF&0.1130&0.0582&0.1385&0.0785&0.1277&0.0721\cr 
    PKEF+DFME & \textbf{0.1320} & \textbf{0.0701} & \textbf{0.1425} & \textbf{0.0794} & \textbf{0.1419} & \textbf{0.0810}\cr
    \bottomrule
    \end{tabular}}
    \end{threeparttable}
    \vspace{-3mm}
    \label{tab:compatible_mtl}
\end{table}

Our proposed DFME can serve as a general module applicable to most existing multi-behavior methods, and we validate this through a compatibility analysis. Specifically, we select some representative multi-behavior methods with MTL, like CRGCN, CIGF and PKEF, as well as multi-behavior methods without MTL, like LightGCN$_M$(LightGCN enhanced with the multi-behavioral graph) and MB-CGCN. Then, we replace their prediction modules with DFME, and compare them with the corresponding original models. The results are shown in Table \ref{tab:compatible_mtl}. As we can see, our proposed DFME improves the performance of all original models. The original LightGCN$_M$ and MB-CGCN benefit significantly from DFME due to their lack of MTL modules. Among multi-behavior methods with MTL, CRGCN exhibits more significant improvement. This is likely due to its original MTL module being relatively basic, which allows for greater compatibility. In contrast, CIGF and PKEF already have sizable MTL modules, resulting in a less pronounced improvement. Overall, the results fully demonstrate the wide compatibility and general applicability of our proposed DFME, which can be integrated into the multi-behavior methods to improve their performance. 

\subsection{Parameter Analysis}
\label{hyper}
\subsubsection{Impact of the number of layers}

\begin{figure}[t]
	\setlength{\belowcaptionskip}{0cm}
	\setlength{\abovecaptionskip}{0cm}
	\subfigure{
        \begin{minipage}[t]{0.47\linewidth}
        \centering
		\label{fig:coefficient_beibei} 
		\includegraphics[width=\textwidth]{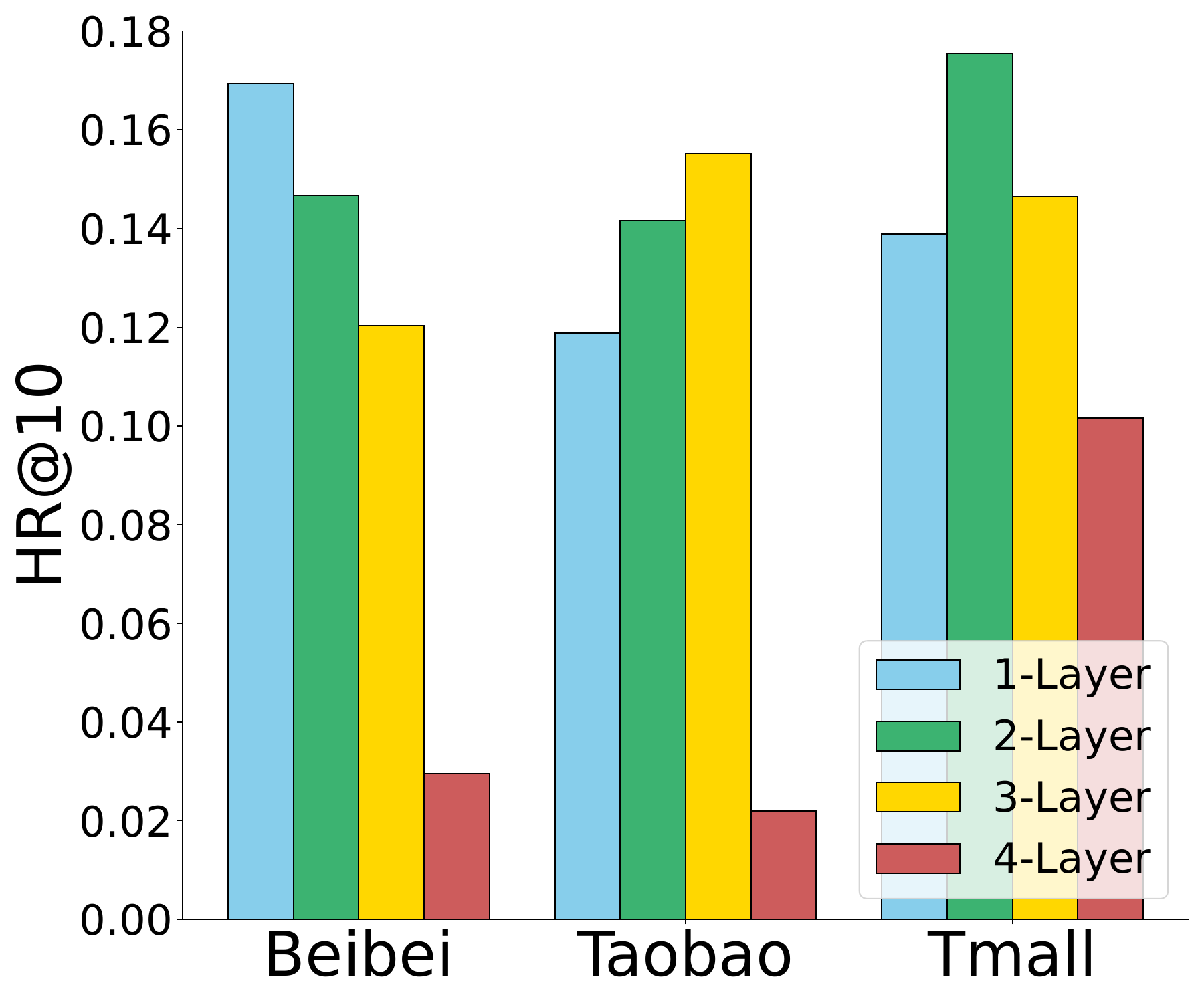}
        \end{minipage}}
	\subfigure{
        \begin{minipage}[t]{0.47\linewidth}
        \centering
		\label{fig:coefficient_taobao} 
		\includegraphics[width=\textwidth]{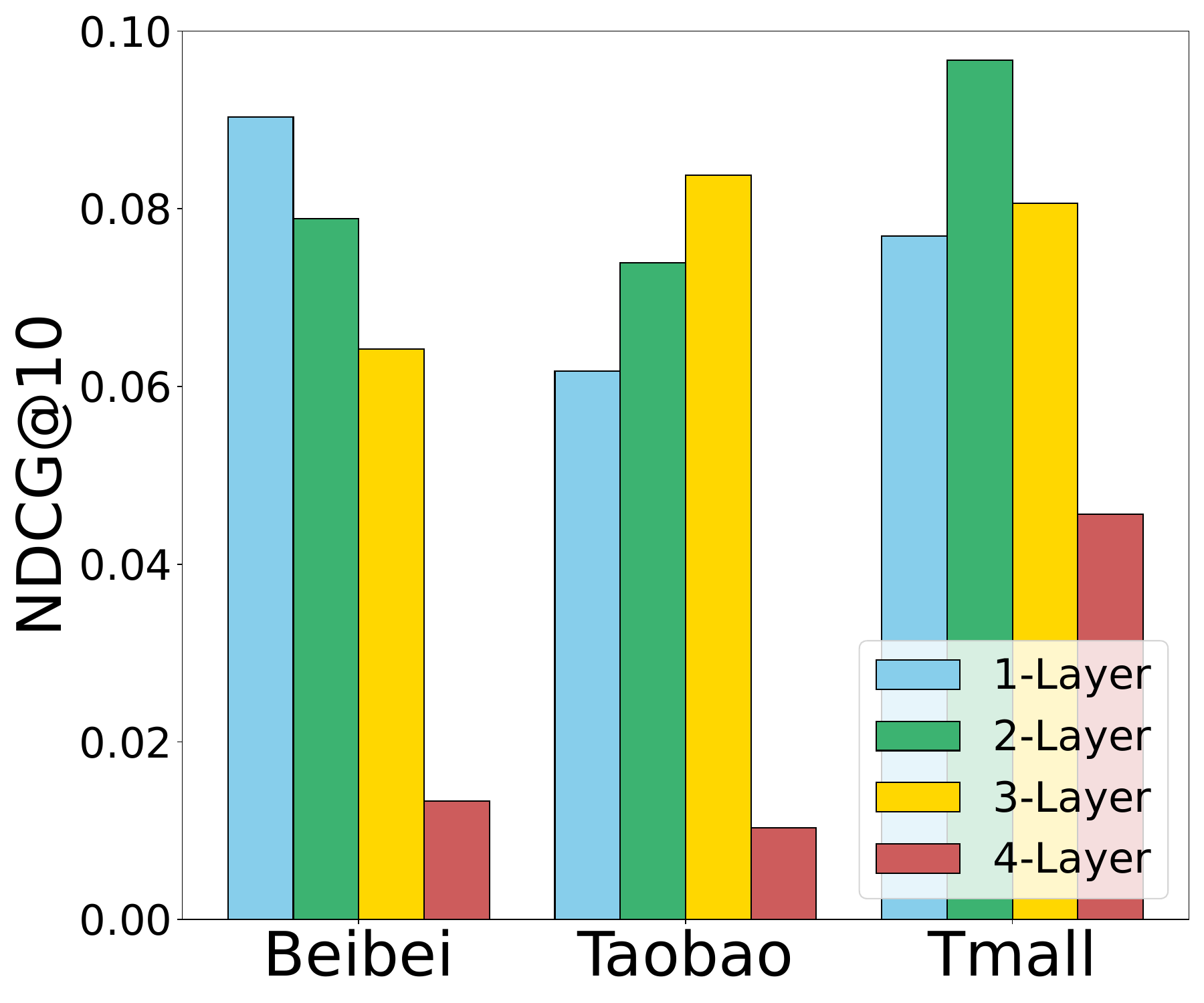}
        \end{minipage}}
	\caption{Impact of GCN layers.}
        \vspace{-3mm}
	\label{fig:layer}
\end{figure}

We investigate the impact of high-order interaction information on model performance by varying the number of GCN layers within the range of \{1, 2, 3, 4\}. As shown in Figure \ref{fig:layer}, the optimal number of layers varies by dataset, which is determined by the relative amounts of noise and useful signals in the high-order information. However, when the number of layers exceeds three, performance significantly declines due to the over-smoothing problem of GCN.

\subsubsection{Impact of temperature hyperparameter}

\begin{figure}[t]
	\setlength{\belowcaptionskip}{0cm}
	\setlength{\abovecaptionskip}{0cm}
	\subfigure{
        \begin{minipage}[t]{0.47\linewidth}
        \centering
		\label{fig:layer_beibei} 
		\includegraphics[width=\textwidth]{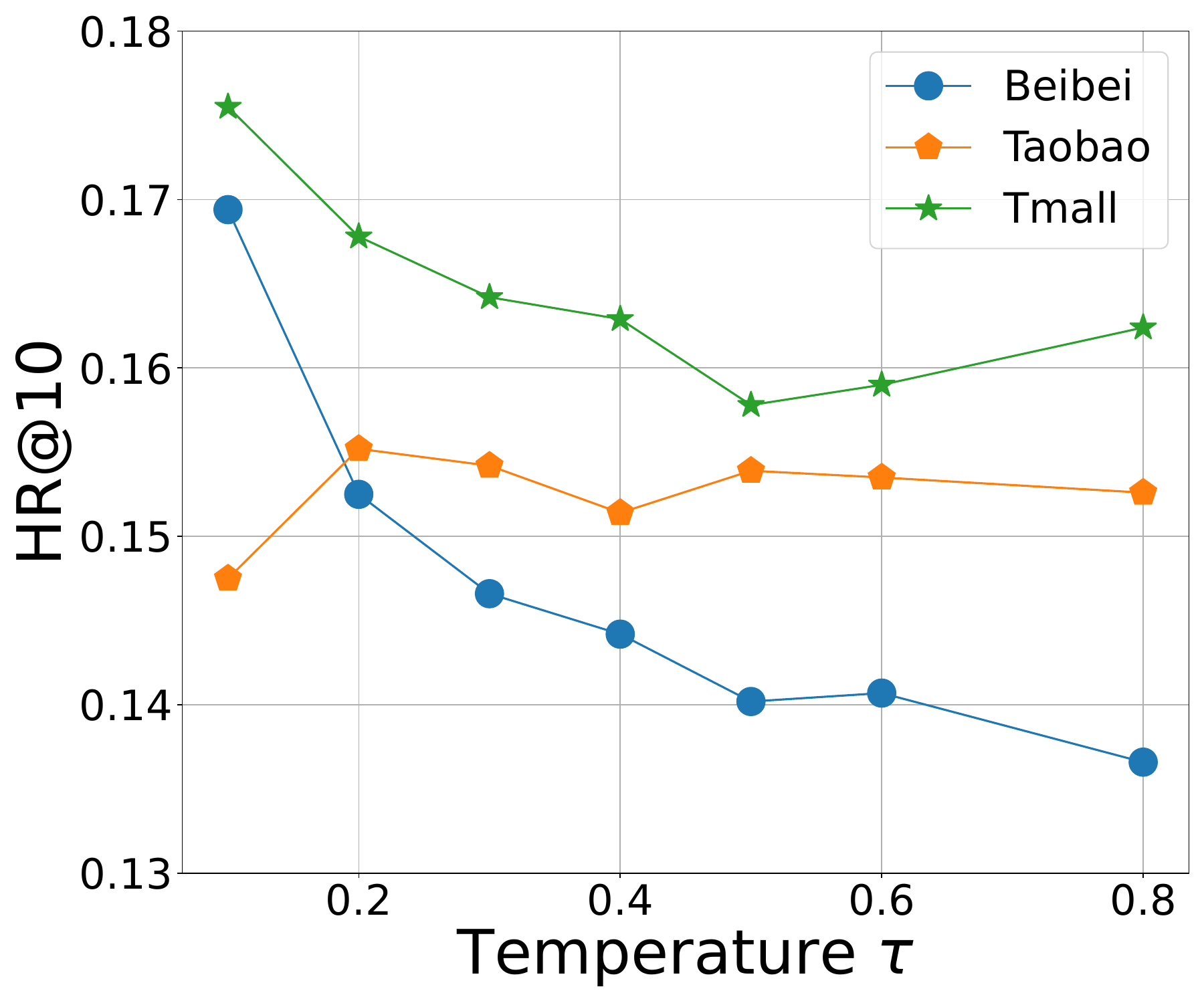}
        \end{minipage}}
	\subfigure{
        \begin{minipage}[t]{0.47\linewidth}
        \centering
		\label{fig:layer_taobao} 
		\includegraphics[width=\textwidth]{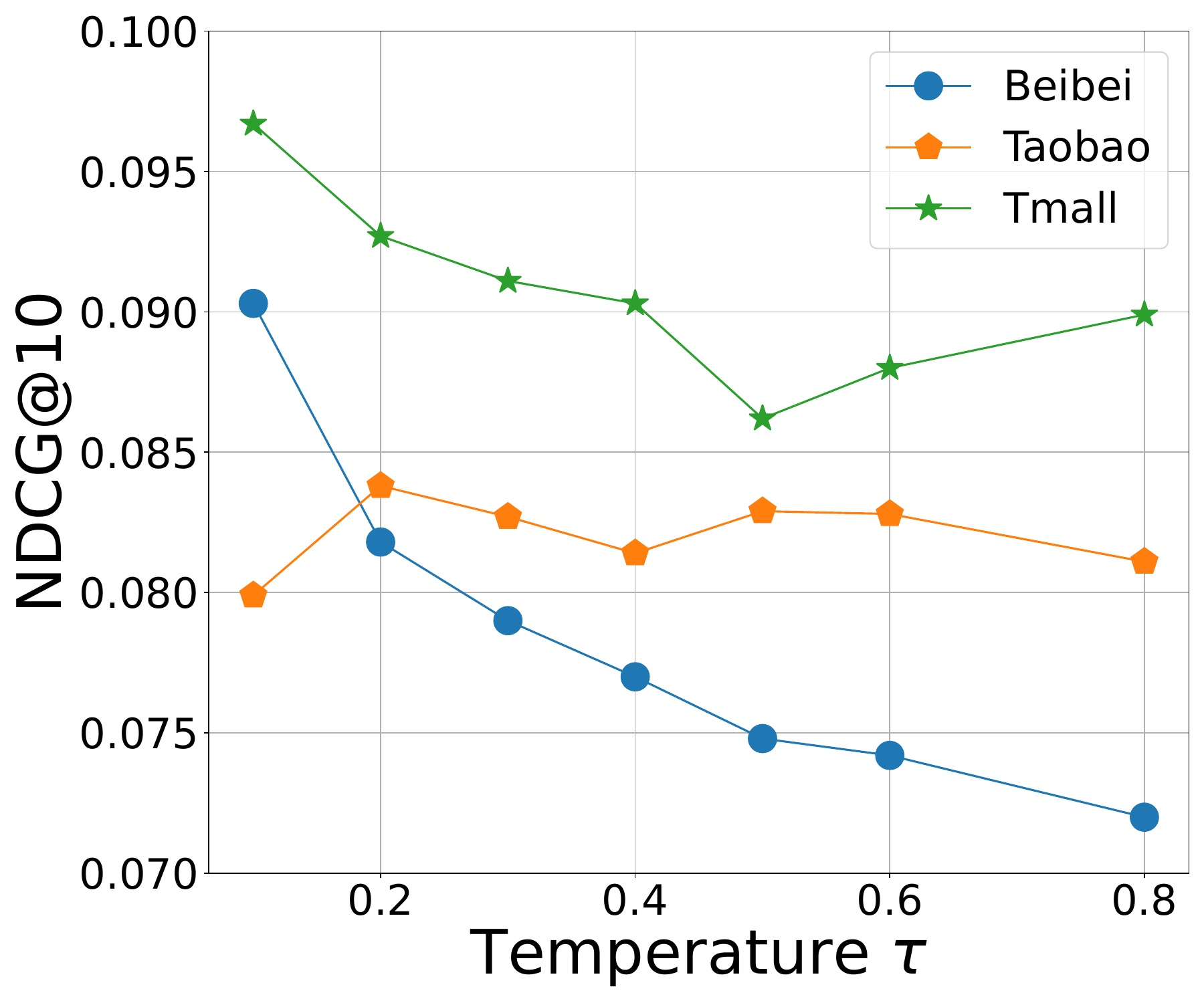}
        \end{minipage}}
	\caption{Impact of temperature hyperparameter.}
        \vspace{-3mm}
	\label{fig:temperature}
\end{figure}

We adjust the temperature hyperparameter in contrastive learning within the range of \{0.1, 0.2, 0.3, 0.4, 0.5, 0.6, 0.8\} and plot the resulting curves (shown in Figure \ref{fig:temperature}). For all three datasets, excessively large temperature coefficients lead to poorer performance, which indicates that a large temperature value will reduce the ability to distinguish negative samples. Additionally, in the Taobao dataset, performance also declines when the temperature coefficient is too small (e.g., 0.1). This may be due to the imbalanced contribution of the samples.

\subsubsection{Impact of the scaling size}

\begin{figure}[t]
	\setlength{\belowcaptionskip}{0cm}
	\setlength{\abovecaptionskip}{0cm}
	\subfigure{
        \begin{minipage}[t]{0.47\linewidth}
        \centering
		\label{fig:scaling_hr} 
		\includegraphics[width=\textwidth]{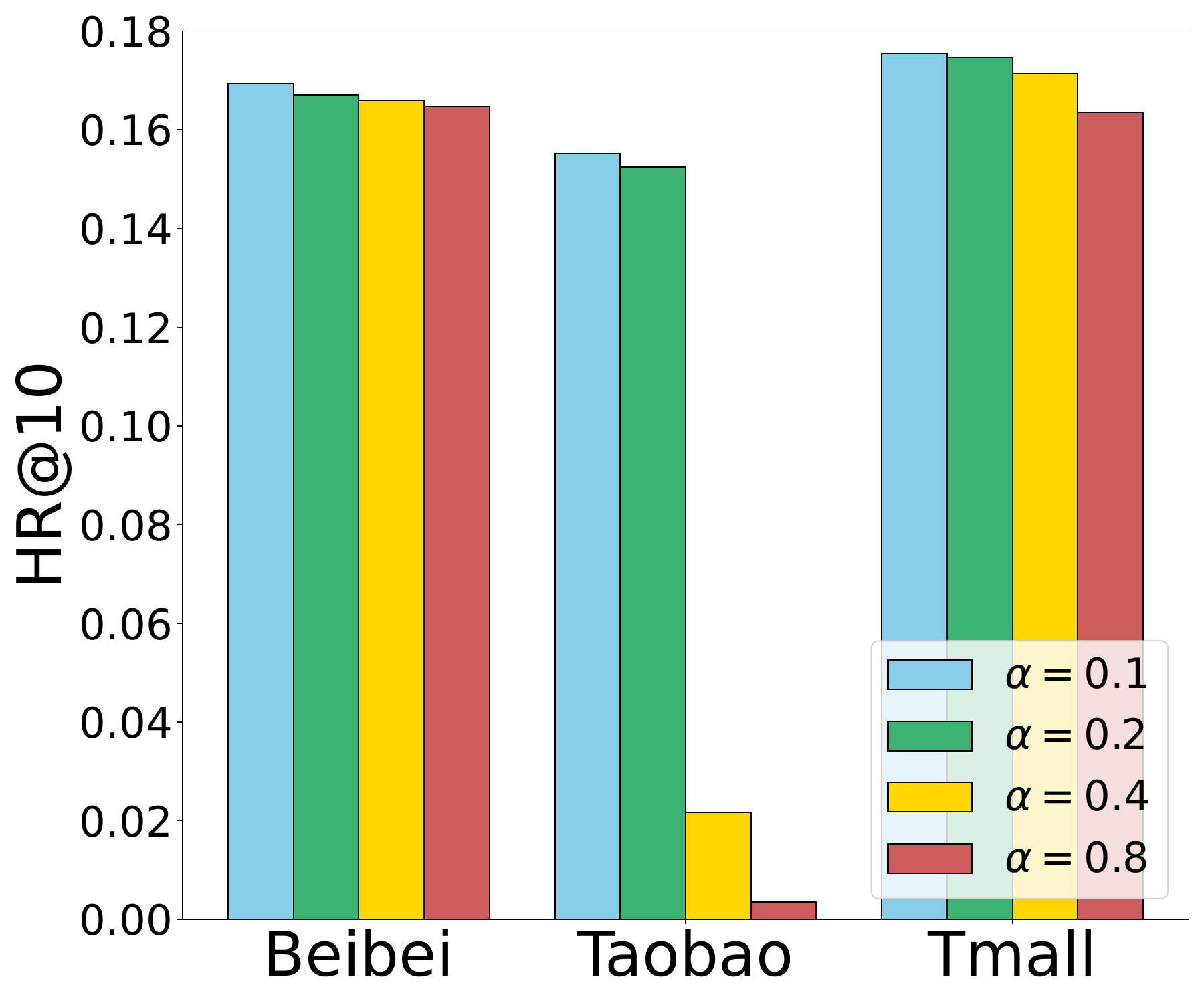}
        \end{minipage}}
	\subfigure{
        \begin{minipage}[t]{0.47\linewidth}
        \centering
		\label{fig:scaling_ndcg} 
		\includegraphics[width=\textwidth]{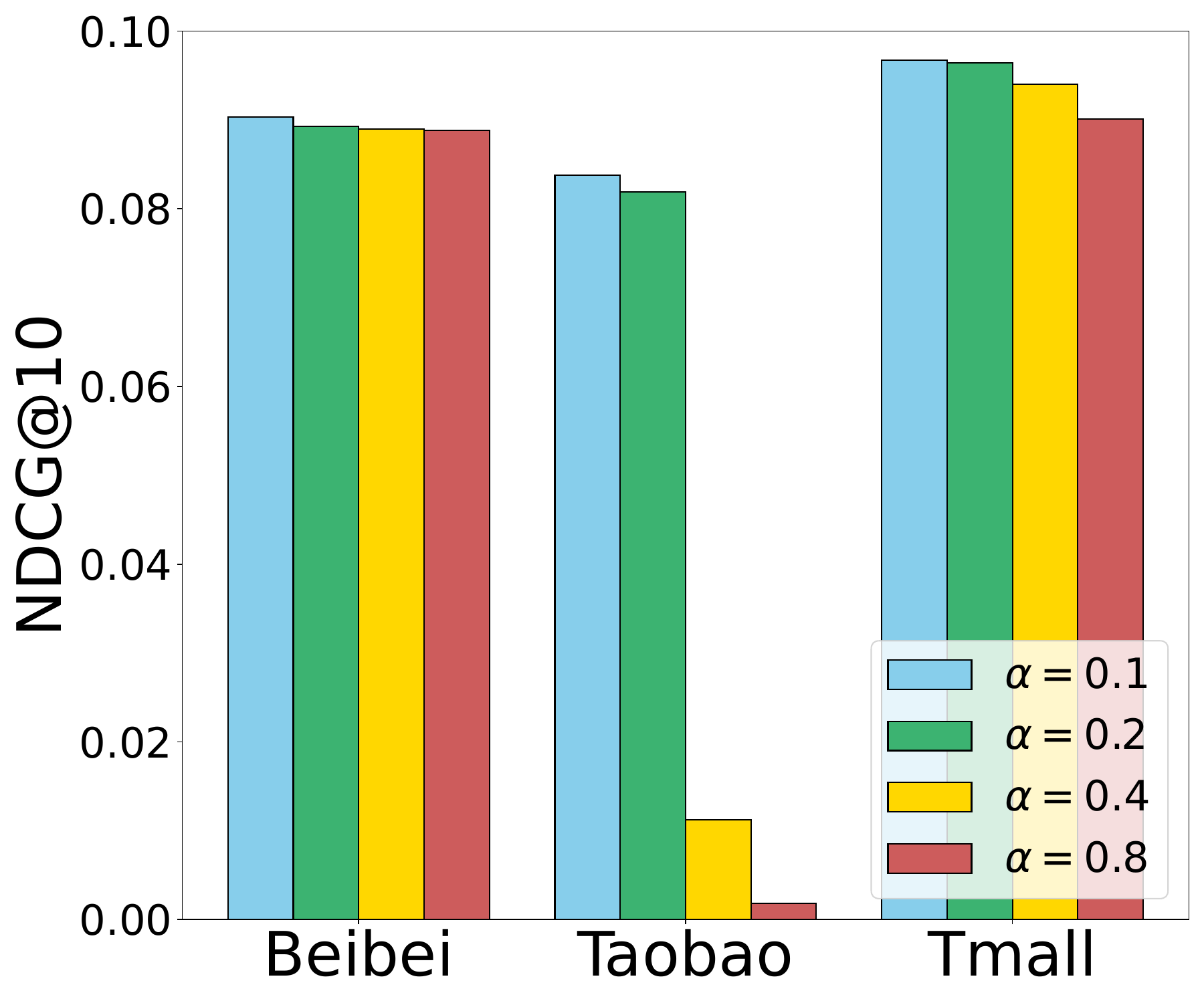}
        \end{minipage}}
	\caption{Impact of the scaling size.}
        \vspace{-3mm}
	\label{fig:scaling}
\end{figure}

With $\beta$ fixed at 0.001, we investigate the impact of the scaling factor $\alpha$ on the forward propagation in multi-task learning. Figure \ref{fig:scaling} shows the results of our search in the range of \{0.1, 0.2, 0.4, 0.8\}. Across all datasets, COPF achieves the highest performance when the scaling factor is small (i.e., 0.1). Besides, the performance gradually decreases as the scaling factor increases, which is particularly evident in the Taobao dataset. These observations highlight the effectiveness of "fine-tuning" the representation space.

\section{CONCLUSION}
\label{future works}
In this paper, we propose the Combinatorial Optimization Perspective based Framework  (COPF) for multi-behavior recommendation. To avoid incorrect modeling of user behavior patterns due to limited perspectives, we propose the Combinatorial Optimization Graph Convolutional Network (COGCN), which treats multi-behavior fusion as a combinatorial optimization problem. COGCN imposes different constraints at various stages of each behavior to restrict the solution space, thereby effectively facilitating the multi-behavior fusion process. To better coordinate the correlations between tasks in MTL methods, we design the Distribution Fitting Expert Network (DFME), which improves both forward and backward propagation. By reducing the distribution differences in features and labels between tasks, DFME alleviates negative information transfer caused by uncoordinated task relationships. We conduct comprehensive experiments on three real-world datasets to verify the superiority of our proposed COPF. Further analysis demonstrates the rationality and effectiveness of the designed COGCN and DFME modules.

\begin{acks}
This work was supported by the STI 2030-Major Projects under Grant 2021ZD0201404.
\end{acks}

\bibliographystyle{ACM-Reference-Format}
\balance
\bibliography{sample-base}

\appendix

\section{APPENDIX}
\subsection{Complexity Analysis}
\label{complexity}
\subsubsection{Time Complexity.}
The main time consumption of COPF lies in the graph convolution operations. The computational complexity of the both steps is $O\left(L\left|\mathcal{E} \right| d\right)$, where $\left|\mathcal{E} \right|$ represents the number of edges in the bipartite graph $\mathcal{G}$, $L$ denotes the number of GNN layers, and $d$ denotes the embedding size. Since the computational complexity does not introduce parameters outside the GNN module, the time complexity of COPF is comparable to that of existing GNN-based methods.

\subsubsection{Space Complexity.} 
The learnable parameters in COPF mainly come from the embeddings of users and items, resulting in a space complexity of $O\left((|\mathbf{U}|+|\mathbf{V}|)d\right)$, which is similar to existing methods. Additionally, as the dense graphs $\mathcal{G}_{k}$ ($k \in \{1,2,...,K\}$) required for graph convolution operation are pre-converted into sparse behavior-specified matrices, there is no need to pay additional space to store these graphs in the computational process. Overall, The memory usage of the model remains within a manageable range during training.

\subsection{Analysis of Data Distribution}

\begin{figure}[H]
	\centering
	\setlength{\belowcaptionskip}{0cm}
	\setlength{\abovecaptionskip}{0cm}
	\includegraphics[width=0.48\textwidth]{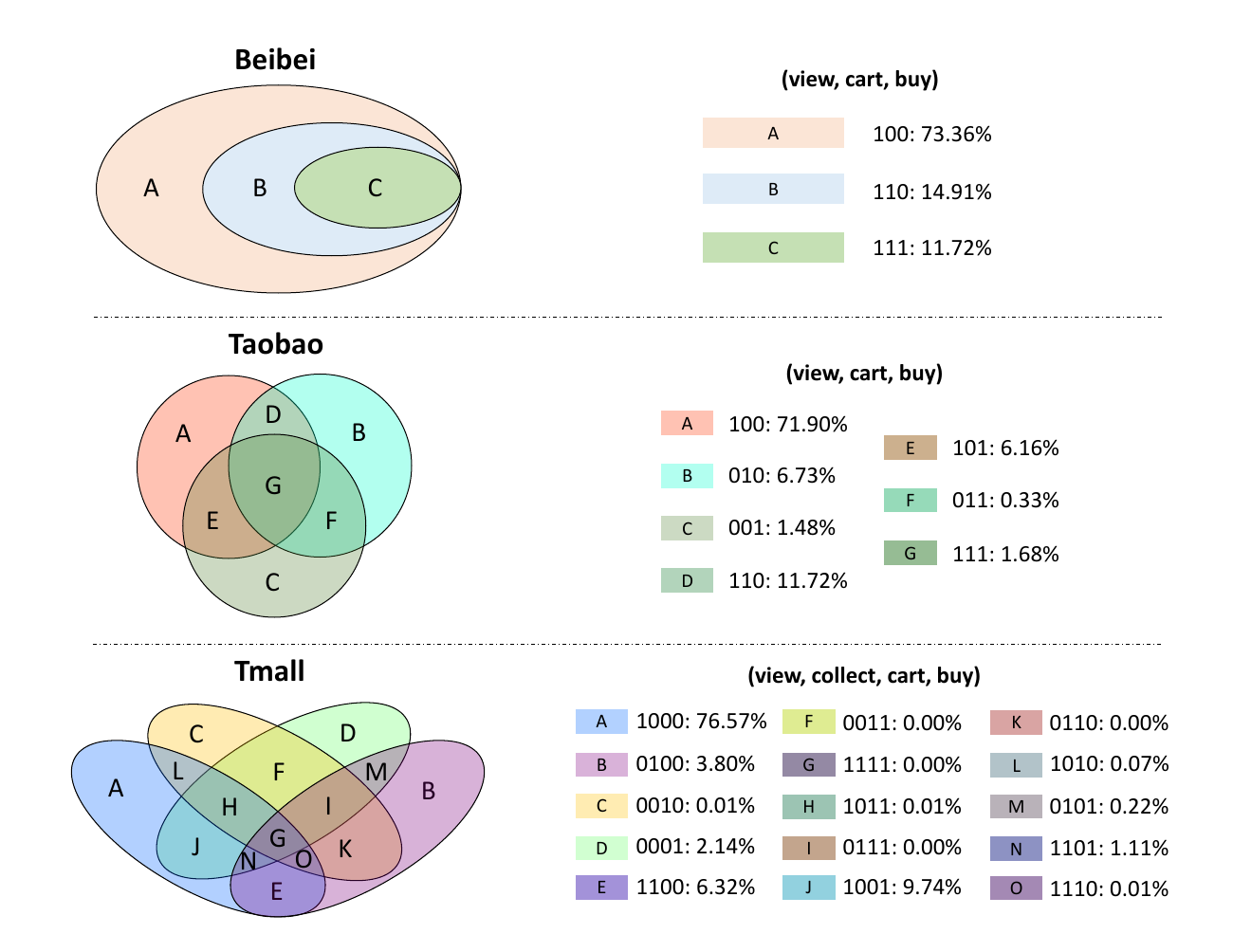}
	\caption{Data distribution of three datasets}
	\label{fig:label_venn}
	\vspace{-4mm}
\end{figure}

Figure \ref{fig:label_venn} shows the data distribution on the three datasets. 1/0 indicates the presence or absence of the corresponding type of behavior. For example, 1001 represents users who only have view and buy behaviors with items. As we can see, there are significant differences in the behavior distribution between datasets, and each dataset contains rich information about user behavior patterns that can be used for learning. For instance, the user behaviors in the Beibei dataset strictly follow the dependency requirements, so the \textbf{COPF-C} variant performs significantly better on Beibei than the other two datasets due to its cascade paradigm(shown in Table \ref{tab:ablation_key}). Moreover, the user behavior patterns are not explicitly available within most datasets since we do not know the specific timing of each user-item interaction. The process of learning user behavior is also implicit. This highlights the necessity of using a general framework(i.e., COGCN) to restrict the solution space by imposing constraints from the perspective of combinatorial optimization to efficiently capture user behavior patterns.

\subsection{Analysis of the Proposed DFME}
\subsubsection{The structure of transfer-based MTL}
\label{classical_mtl}

\begin{figure}[H]
	\centering
	\setlength{\belowcaptionskip}{0cm}
	\setlength{\abovecaptionskip}{0cm}
	\includegraphics[width=0.47\textwidth]{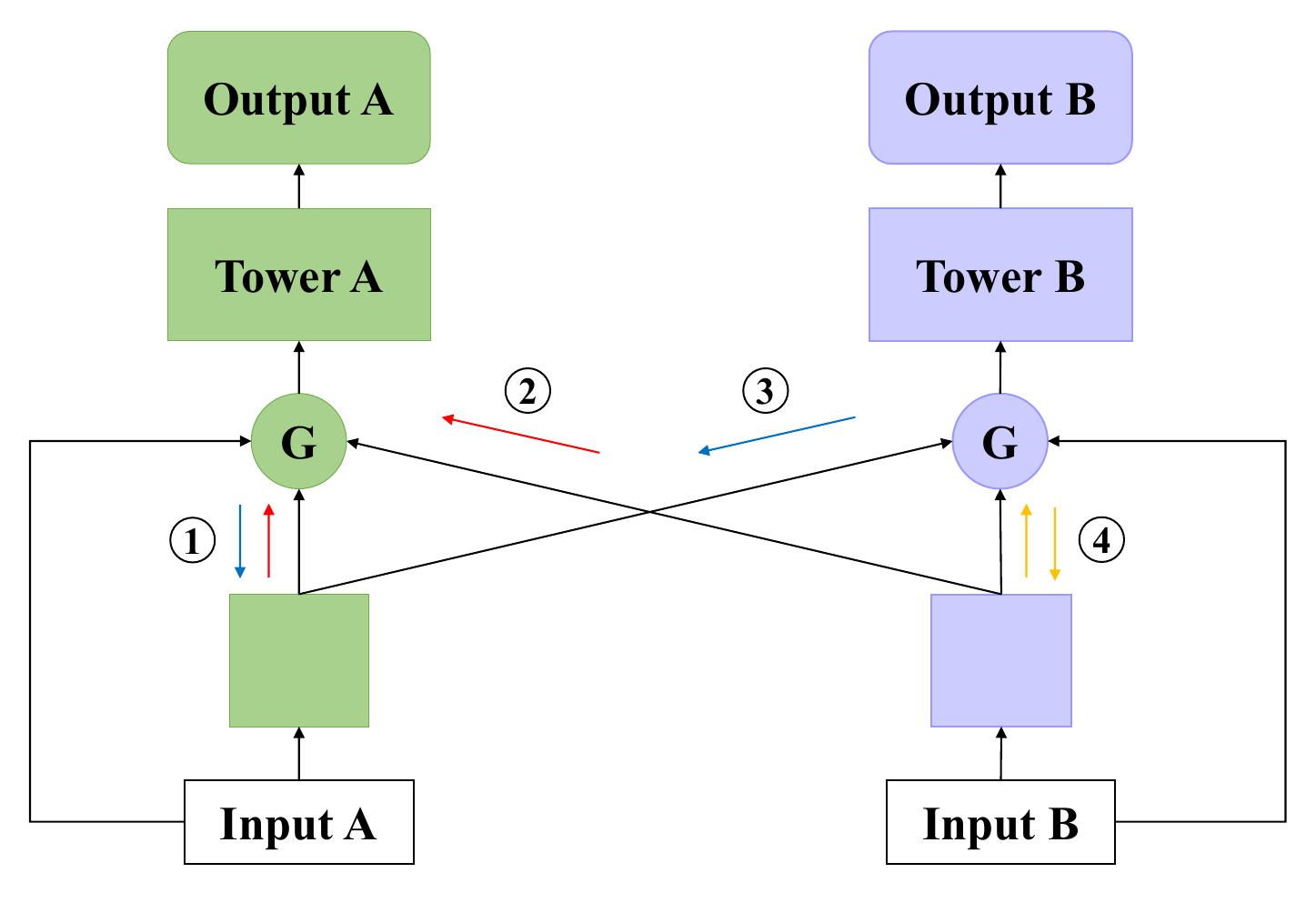}
	\caption{The structure of transfer-based MTL. Red line represents the forward propagation that affects the target task A, blue line denotes the backward propagation that affects the target task A, and yellow line represents the irrelevant process.}
	\label{fig:mtl_structure}
	\vspace{-4mm}
\end{figure}

\begin{figure*}[ht]
	\centering
	\setlength{\belowcaptionskip}{0cm}
	\setlength{\abovecaptionskip}{0cm}
	\includegraphics[width=\textwidth]{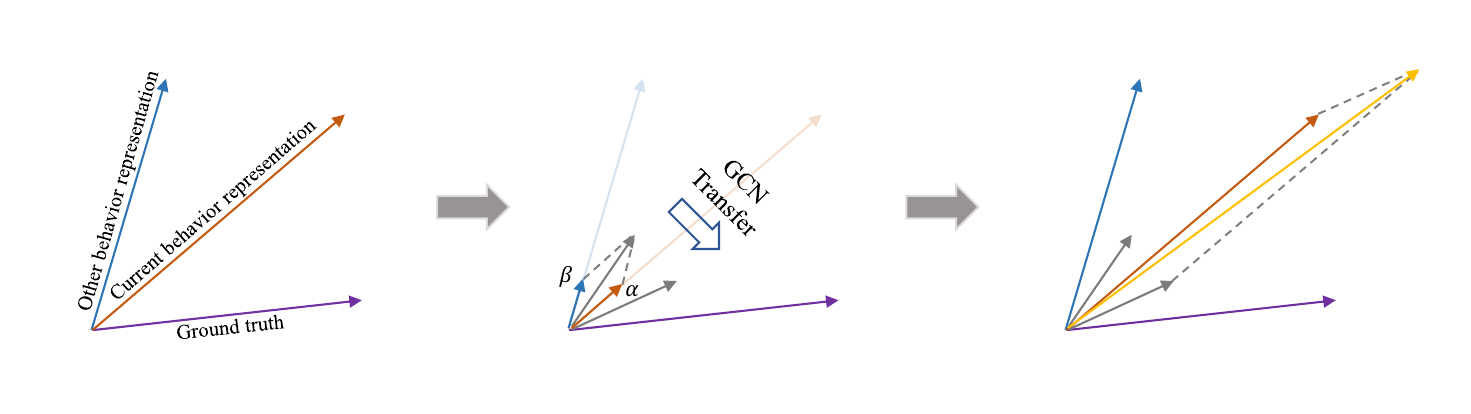}
	\caption{Refine with other behavior representions.}
	\label{fig:engle}
	\vspace{-4mm}
\end{figure*}

A basic transfer-based MTL network is shown in Figure \ref{fig:mtl_structure}. The task corresponding to the target behavior is designated as the target task (Task A), while other tasks are auxiliary tasks(Task B and other tasks). As we can see, there are four paths associated with these two tasks. For target task A, path 1 is the main path for this task, and path 4 becomes an irrelevant path due to the decoupling of task inputs \cite{pkef}. Paths 2 and 3 are information interaction paths resulting from the aggregation between tasks, which are the main correlations that DFME needs to coordinate. Specifically, path 2 affects the target task A at the feature level during forward propagation, and path 3 affects the target task A at the label level during backward propagation.

\subsubsection{Shortcomings of existing optimal MTL}
\label{shortcoming_pkef}
PKEF introduces a projection mechanism during aggregation to disentangle the shared and unique parts for other behavioral experts. The shared part is used for aggregation, while the unique part is used for auxiliary learning. While somewhat effective, this method has two main issues: \textbf{1)} For instance, with the behaviors "\textit{cart}" and "\textit{buy}", PKEF assumes that the shared part is the information related to both behaviors occurring together (e.g., \textit{cart \& buy}), while the unique part is information related to only a single behavior (e.g., \textit{only cart}), thus further sets an auxiliary loss for the unique part accordingly (e.g., labels for \textit{cart w/o buy}). In reality, "\textit{not buy}" is also valuable information for learning behavior \textit{buy}, which should exist in the shared part, and the unique part should be completely irrelevant information. This incorrect projection mechanism can compromise the effectiveness of the information during aggregation. Additionally, the projection mechanism ensures that the representation space of the auxiliary behavior remains consistent with that of the current behavior during aggregation. However, since the model is still in the training phase, the representation space for the target behavior is unstable and may be inaccurate. Enforcing such alignment could introduce detrimental information, which can potentially prevent the model from converging to the optimal distribution and adversely affect its generalization performance. \textbf{2)} PKEF overlooks the impact of gradient coupling during aggregation, where gradient updates from auxiliary tasks affect the target task. The above two problems may cause the model's inability to accurately fit the target behavior distribution, leading to negative transfer problem. 

\subsubsection{Method of the proposed DFME}
\label{method_dfme}
Our proposed MTL network, named DFME, coordinates the relationship between target and auxiliary tasks in two key aspects to control negative information transfer. Specifically, in forward propagation, contrastive learning between tasks is utilized to enable the target behavior to obtain effective information from the auxiliary behaviors, thereby reducing the distribution gap between the target and auxiliary behaviors when generating behavior-specific experts. Meanwhile, we spatially adapt the representation of the current behavior to fit the aggregation process by generating behavior-fitting experts for each task, thereby preventing interference among different task behaviors. As shown in Figure \ref{fig:engle}, for the current behavior $k$ and one of the other behaviors $k^{\prime}$, we fuse the representations of behavior $k$ and $k^{\prime}$ after multiplying them by small weight coefficients respectively. This yields a fitted representation that is appropriately sized and slightly different in direction from the representation of behavior $k^{\prime}$. We then obtain the final task output representation by capturing the effective information about behavior $k^{\prime}$ contained in this fitted representation through a graph convolutional network. Finally, we use behavior-specific expert for the current behavior and behavior-fitting experts for other behaviors during aggregation. In summary, the above steps can be outlined as follows: the effective information contained in the interaction between behaviors is used to refine the representation space of the current behavior while ensuring that the generalization performance of the model is not affected.

During backward propagation, for a transfer-based MTL using decoupled input ($\mathbf{e}_{u}^{k^{\prime}}$,$\mathbf{e}_{v}^{k^{\prime}}$) for each task $k^{\prime} \in \{1,2,...,K\}$ , the aggregation process of the auxiliary task $k$ is:
\begin{displaymath}
    \mathbf{o}^{k} = \sum_{k^{\prime}=1}^{K}\mathbf{g}^k(k^{\prime})\cdot\mathbf{e}^{k^{\prime}}
\end{displaymath}
where $\mathbf{e}^{k^{\prime}}$ is the output of the expert, $\mathbf{g}^k(k^{\prime})$ indicates the $k^{\prime}$-th element of the gating vector $\mathbf{g}^k$. Due to the introduction of behavior-fitting experts (i.e., $f_k(\cdot)$) in the forward propagation to refine the representation space for each behavior, the aggregation process can be further represented as:
\begin{displaymath}
    \mathbf{o}^k=\sum_{k^{\prime}=1,k^{\prime}\neq k}^K{\mathbf{g}^k(k^{\prime})\cdot\mathbf{e}^{k, k^{\prime}}}+\mathbf{g}^k(k)\cdot\mathbf{e}^{k}
\end{displaymath}
where $\mathbf{e}^{k, k^{\prime}}=f_k(\mathbf{e}_{u}^{k},\mathbf{e}_{v}^{k},\mathbf{e}_{u}^{k^{\prime}},\mathbf{e}_{v}^{k^{\prime}})$ denotes the output of the behavior-fitting expert. And the loss function for task $k$ can be defined as: 
\begin{displaymath}
\mathcal{L}_{k}=L(h^{k}(\mathbf{o}^{k})-\hat{o}_{uv}^{k})
\end{displaymath}
where $h^{k}(\cdot)$ is the tower function, $L(\cdot)$ denotes the loss function. Then we can obtain the gradient of the auxiliary task $k$ with respect to the behavior target behavior $t$ as follows: 
\begin{displaymath}
\begin{aligned}
{\frac{\partial \mathcal{L}_{k}}{\partial \mathbf{e}_{*}^t}}
&=\frac{\partial h^k(\mathbf{g}^k(t)\cdot \mathbf{e}^{k, t})}{\partial \mathbf{e}_{*}^t}*L^{\prime}(h^{k}(\mathbf{o}^k)-\hat{o}_{uv}^k) \\
&=\frac{\partial h^k(\mathbf{g}^k(t)\cdot f_k(\mathbf{e}_{u}^{k},\mathbf{e}_{v}^{k},\mathbf{e}_{u}^{t},\mathbf{e}_{v}^{t}))}{\partial \mathbf{e}_{*}^t}*L^{\prime}(h^{k}(\mathbf{o}^k)-\hat{o}_{uv}^k)
\end{aligned}
\end{displaymath}
where $\mathbf{e}_{*}^t \in \{\mathbf{e}_{u}^{t},\mathbf{e}_{v}^{t}\}$. It can be seen that the gradient update of the auxiliary task still affects the target behavior. Therefore, we stop the gradient updates from the auxiliary loss to the target behavior in order to alleviate the potential negative transfer caused by gradient coupling in the multi-behavior prediction step. 









\end{document}